\documentclass[pra,twocolumn,showpacs,letterpaper,showpacs,superscriptaddress]{revtex4}
\usepackage{graphicx,amsmath,amssymb,amsfonts,latexsym,color,dcolumn,bm,epsfig,subfigure,pstricks-add}
\newrgbcolor{cinza1}{0.5 0.5 0.5}
\newrgbcolor{cinza2}{0.6 0.6 0.6}
\newrgbcolor{cinza3}{0.7 0.7 0.7}
\newrgbcolor{cinza4}{0.8 0.8 0.8}
\newrgbcolor{cinza5}{0.9 0.9 0.9}

\begin{document}

\title{Casimir Interactions for Anisotropic Magnetodielectric Metamaterials}

\author{F. S. S. Rosa}
\affiliation{Theoretical Division, Los Alamos National Laboratory, Los Alamos, NM 87545, USA}

\author{D. A. R. Dalvit}
\affiliation{Theoretical Division, Los Alamos National Laboratory, Los Alamos, NM 87545, USA}

\author{P. W. Milonni}
\affiliation{Theoretical Division, Los Alamos National Laboratory, Los Alamos, NM 87545, USA}

\date{\today}

\begin{abstract}
We extend our previous work [Phys. Rev. Lett. {\bf 100}, 183602 (2008)] on the generalization of the Casimir-Lifshitz theory to treat anisotropic magnetodielectric media, focusing on the forces between metals and magnetodielectric metamaterials and on the possibility of inferring magnetic effects by measurements of these forces. We present results for metamaterials including structures with uniaxial and biaxial magnetodielectric anisotropies, as well as for structures with isolated metallic or dielectric properties that we describe in terms of filling factors and a Maxwell Garnett approximation. The elimination or reduction of Casimir ``stiction" by appropriate engineering of metallic-based metamaterials, or the indirect detection of magnetic contributions, appear from the examples considered to be very challenging, as small background Drude contributions to the permittivity act to enhance attraction
over repulsion, as does magnetic dissipation. In dielectric-based metamaterials the magnetic
properties of polaritonic crystals, for instance, appear to be too weak for repulsion to overcome attraction. We also discuss Casimir-Polder
experiments, that might provide another possibility for the detection of magnetic effects.
\end{abstract}


\maketitle

\section{Introduction}

The last decade has witnessed a huge activity in the development of metamaterials (MMs) 
\cite{reviewsMM}, boosted by the possibility that such engineered media may give rise to novel optical properties at selected frequency ranges, including negative refraction \cite{negative_refraction}, perfect lensing 
\cite{perfect_lens}, and cloaking \cite{cloaking}, among others. Such striking phenomena, inaccessible with natural materials, are all possible due to the significant magnetic activity built into metamaterials, starting at microwave frequencies and going all the way up to the optical range. Generally speaking, metamaterials are made of micro- and nanostructures carefully designed to collectively endow them with a particular electromagnetic property. It is generally
desirable that these structures should be smaller than the wavelength of the incident radiation, so that they are seen by the incoming waves as artificial {\lq\lq atoms\rq\rq}.  This fact often allows the use of an effective medium approximation to describe metamaterials in terms of an effective electric permittivitiy tensor $\boldsymbol{\epsilon}(\omega)$ and an effective magnetic permeability tensor $\boldsymbol{\mu}(\omega)$, which incorporate the typical optical anisotropy of metamaterials. 

Recent years have also witnessed an increased interest in Casimir physics \cite{Casimir,reviewsCasimir} thanks to improved
precision measurements \cite{casimirexperiments} of the force between material objects separated by micron and sub-micron gaps. Quantum vacuum fluctuations are modified by the presence of material boundaries, and this typically results in an attractive Casimir force that depends sensitively on the shape and the optical properties of the boundaries. While the Casimir force offers new possibilities for nanotechnology, such as actuation mediated by the quantum vacuum, it also
presents some challenges, as micro- and nanoelectromechanical systems (MEMS and NEMS) may stick together and cease to work due to the attractive nature of van der Waals and Casimir forces. A strongly suppressed Casimir attraction, or even repulsive Casimir forces, would provide an anti-``stiction" effect. 
Repulsive Casimir forces between two objects 1 and 2, immersed in a background medium 3, may come in a variety of ways. One possibility involves non-magnetic media only, for which repulsion
happens when the electric permittivities evaluated at imaginary frequencies satisfy the relation 
$\epsilon_1(i \omega) < \epsilon_3(i \omega) < \epsilon_2(i \omega)$ \cite{dielectricrepulsion}. Another 
possibility, first predicted by Boyer \cite{boyer}, involves magnetodielectric media: there is a repulsive force
when a perfectly conducting plate is placed near a perfectly permeable one with vacuum in between. Some years later it was shown that Casimir repulsion can also occur between real ({\it i.e.}, non-ideal)  magnetodielectric media, as long one medium is mainly electric and the other one is mainly magnetic \cite{klich}. However, this possibility has been considered unphysical \cite{capassocomment}, as naturally occurring materials, even strong magnets at low frequencies \cite{ferrites}, do not show significant magnetic response at near-infrared and optical frequencies, which has been assumed as a prerequisite for repulsion between Casimir plates separated by typical experimentally relevant distances of $d=0.1-1\mu{\rm m}$. On the other hand, recent developments in nanofabrication have resulted in metamaterials with magnetic response in the visible range of the electromagnetic spectrum
\cite{grigorenko, shalaev, dolling}, fueling the hope for Casimir repulsion \cite{henkel,leonhardt,irina,zhu}.
The expectation is that, by tuning this magnetic response to the right frequency range and making it strong enough, one could produce an experimentally measurable Casimir repulsion between, say, a MM slab and a thin metallic plate, or at least a significantly reduced attraction. 

Unfortunately, this is easier said than done. The major issue is that the Casimir force between real dispersive materials is a broadband frequency phenomenon, as shown by the Lifshitz formula expressing the force between two semispaces as an integral over all (imaginary) frequencies with an exponential cut-off $c/d$ \cite{lifshitz}. For Casimir repulsion purposes, this requires a magnetic response strong enough to dominate the electric response of the material in a broad range of frequencies, which typically is not the case for the magnetic resonances present in metamaterials. 
In addition, several metamaterials have metallic inclusions that produce a low-frequency Drude-like behavior in $\boldsymbol{\epsilon}(\omega)$,
whose contribution to the Liftshitz formula dominates over any possible magnetic response that the metamaterial may have, 
making attractive a Casimir force that would otherwise be predicted to be repulsive. 

We have recently addressed many of these issues in the context of the Casimir-Lishitz theory and metamaterials \cite{usPRL}.
The purpose of the present work is to further investigate the physics of Casimir interactions between metamaterials, focusing on effects not previously considered in depth in the Casimir literature, such as optical anisotropy in magnetodielectrics and the feasibility of the crossover from attractive to repulsive Casimir forces with realistic metallic-based and dielectric-based metamaterials.    
%%%%%%%%%%%%%%%%%%%%%%%%%%%%%%%%%%%%%%

\section{Casimir-Lifshitz force between anisotropic magnetodielectric materials}

\subsection{The scattering approach}

Techniques for the evaluation of the Casimir force have evolved very quickly in the last few years, paving the way for precise analytical \cite{analyticalsolutions} and numerical \cite{numericalsolutions} calculations in non-trivial geometries. A particularly appealing method is the so-called scattering approach, pioneered in Casimir physics by Balian and Duplantier to compute the free energy of the electromagnetic field in regions bounded by material boundaries of arbitrary smooth shape \cite{BalDup}.
The free energy is expressed as a convergent multiple scattering expansion of ray trajectories propagating between the material
boundaries. This method, first used for perfect conductors,  was extended to real materials in recent works \cite{french,MITgroup}, allowing in principle the computation of the Casimir interaction between arbitrarily shaped material scatterers. 

Since a thorough discussion of the scattering approach would take us too far afield, we simply present the formula for the zero-temperature Casimir energy per unit area $A$ between two parallel plates separated by a vacuum gap of width $d$:
\begin{eqnarray}
\label{energyscatteringapproach}
\frac{E(d)}{A} = \hbar \int_0^{\infty} \frac{d\xi}{2\pi} \log \det \left[ 1 - {\cal R}_1 e^{-{\cal K} d}{\cal R}_2 e^{-{\cal K} d}\right] ,
\end{eqnarray} 
where ${\cal R}_j = {\cal R}_j ({\bf k_{\|}},{\bf k^{\prime}_{\|}},p,p',\omega=i \xi)$ is the reflection operator associated with reflection on the $j$-th plate ($j=1,2$). Here ${\bf k_{\|}}$ and ${\bf k^{\prime}_{\|}}$ are the transverse wave vectors ({\it i.e}, projected onto the planar interfaces) for incident and reflected waves, respectively, and $p$ and $p'$ are their respective polarizations (transverse electric (TE) or transverse magnetic (TM)).
The operator $\exp(- {\cal K} d)$ represents one-way propagation between the two plates,
and has matrix elements
\begin{eqnarray}
\label{eq2}
\langle {\bf k}_{\|} | e^{- {\cal K} d}  | {\bf k}_{\|}^{\prime} \rangle = \exp(- d \sqrt{k^2_{\|} + \xi^2/c^2}) \; \delta^{(2)}({\bf k}_{\|} - {\bf k}^{\prime}_{\|}) .
\end{eqnarray}
When both plates present homogeneity in the plane of the interface, only specular reflection takes place,
and ${\cal R}_j$ is also diagonal in the transverse momentum basis. This means that
\begin{eqnarray}
\label{eq3}
\langle {\bf k}_{\|} | {\cal R}_j | {\bf k}_{\|}^{\prime} \rangle  = {\bf R}_j \; \delta^{(2)}({\bf k}_{\|} - {\bf k}^{\prime}_{\|}) , 
\end{eqnarray}
where ${\bf R}_j$ is the  $2\times 2$ reflection matrix on 
the $j$-th plate.
Note that the reflection matrices here are evaluated at imaginary frequencies $\omega=i \xi$,
and this requires the well-known analytic properties of the permittivities and permeabilities
in the complex frequency plane.
For general anisotropic media these reflection matrices are defined as
\begin{eqnarray}
\label{ReflectionMatrices}
{\bf R}_j = \left[
\begin{array}{cc}
   r^{{\rm TE, TE}}_j (i \xi, {\bf k}_{\|}) &  r^{{\rm TE, TM}}_j (i \xi, {\bf k}_{\|}) \\
   r^{{\rm TM, TE}}_j (i \xi, {\bf k}_{\|}) &  r^{{\rm TM, TM}}_j (i \xi, {\bf k}_{\|}) 
\end{array} \right] ,
\end{eqnarray}
where $r^{p,p'}_j$ is the ratio of the amplitudes of a reflected field with $p'$-polarization and an incoming field with $p$-polarization. 

Using Eqs. (\ref{eq3}) and (\ref{eq2}) in Eq. (\ref{energyscatteringapproach}), we get after some manipulations
\begin{equation}
\label{EnergyAnisotropic}
\frac{E(d)}{A} = \hbar \int_0^{\infty} \hspace{-2pt} \frac{d\xi}{2\pi} \int \frac{d^2 {\bf k}_{\|}}{(2\pi)^2} \log \det \left[1 - {\bf R}_1 \cdot {\bf R}_2 e^{-2 K_3 d}\right] ,
\end{equation}
and the expression for the force per unit area follows:
\begin{equation}
\label{eq1}
\frac{F(d)}{A}= 2 \hbar \int_0^{\infty} \hspace{-2pt}  \frac{d\xi}{2 \pi} 
\int  \frac{d^2 {\bf k}_{\|}}{(2 \pi)^2}
\, K_3 \, {\rm Tr} \,
\frac{{\bf R}_1 \cdot {\bf R}_2 \, e^{-2K_3 d} }{
1-{\bf R}_1 \cdot {\bf R}_2 \, e^{-2K_3 d}} ,
\end{equation}
where $K_3=\sqrt{k^2_{\|}+ \xi^2/c^2}$. A positive (negative) value of the force corresponds to
attraction (repulsion). Despite the fact that we have assumed homogeneity on each of the planar interfaces (which is a reasonable assumption when describing metamaterials with an effective medium approach), Eq. (\ref{eq1}) is still fairly general: it may be applied to dispersive, dissipative and anisotropic media; all that is needed are the appropriate reflection matrices.

%
%%%%%%%%%%%%%%%%%%%%%%%%%%%%%%%%%%%%%%%%%%%%%
\begin{figure}
\begin{center}
\scalebox{0.375}{\includegraphics{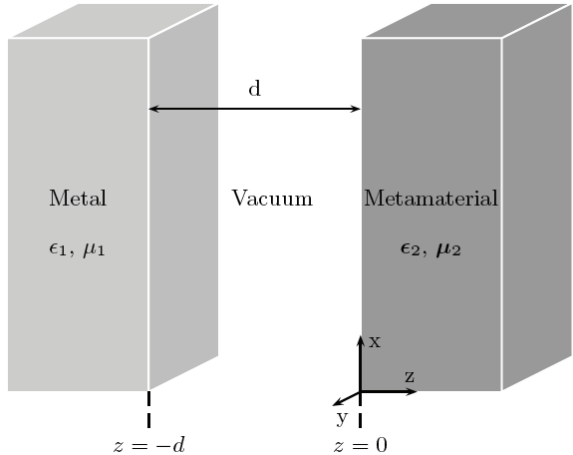}}
\caption{Typical setup used throughout this paper to compute the Casimir-Lifshitz force between a metal and a metamaterial.}
\label{fig1}
\end{center}
\end{figure}
%%%%%%%%%%%%%%%%%%%%%%%%%%%%%%%%%%%%%%%

Let us consider the setup depicted in Fig. 1, where we have a metallic semi-space occupying the region $z<-d$ facing a magnetodielectric semi-space $z>0$. The reflection matrix 
${\bf R}_1$ characterizing the metal-vacuum interface is  given by the standard Fresnel coefficients \cite{BornWolf}  
\begin{eqnarray}
\label{FresnelCoefficients}
&&r^{{\rm TE, TE}}_1(i\xi, {\bf k}_{\|}) = \frac{K_3 - \sqrt{k_{\|}^2 + \epsilon_1(i \xi) \xi^2/c^2}}{K_3 + \sqrt{k_{\|}^2 + \epsilon_1(i \xi) \xi^2/c^2}} , \nonumber \\
&&r^{{\rm TM, TM}}_1(i\xi, {\bf k}_{\|}) = \frac{\epsilon_1 (i \xi) K_3 - \sqrt{k_{\|}^2 + \epsilon_1(i \xi) \xi^2/c^2}}{\epsilon_j(i \xi) K_3 + \sqrt{k_{\|}^2 +  \epsilon_1(i \xi) \xi^2/c^2}} , \nonumber \\
&&r^{{\rm TE, TM}}_1(i\xi, {\bf k}_{\|}) = r^{{\rm TM, TE}}_1(i\xi, {\bf k}_{\|}) = 0 ,
\end{eqnarray}
where $\epsilon_1(\omega)$ is the permittivity of the metal.
The elements of  ${\bf R}_2$ are only given by Fresnel-like formulas when the MM is isotropic, in which case
\begin{eqnarray}
\label{FresnelCoefficientsMM}
&&r^{{\rm TE, TE}}_{2_{\rm iso}}(i\xi, {\bf k}_{\|}) = \frac{\mu_2 (i \xi) K_3 - \sqrt{k_{\|}^2 + \mu_2 (i \xi)\epsilon_2(i \xi) \xi^2/c^2}}{\mu_2 (i \xi)K_3 + \sqrt{k_{\|}^2 + \mu_2 (i \xi)\epsilon_2(i \xi) \xi^2/c^2}} , \nonumber \\
&&r^{{\rm TM, TM}}_{2_{\rm iso}}(i\xi, {\bf k}_{\|}) = \frac{\epsilon_2 (i \xi) K_3 - \sqrt{k_{\|}^2 + \mu_2 (i \xi)\epsilon_2(i \xi) \xi^2/c^2}}{\epsilon_2(i \xi) K_3 + \sqrt{k_{\|}^2 +  \mu_2 (i \xi)\epsilon_2 (i \xi) \xi^2/c^2}} , \nonumber \\
&&r^{{\rm TE, TM}}_{2_{\rm iso}}(i\xi, {\bf k}_{\|}) = r^{{\rm TM, TE}}_{2_{\rm iso}}(i\xi, {\bf k}_{\|}) = 0 ,
\end{eqnarray}
where $\epsilon_2, \mu_2$ are respectively the permittivity and the permeability of the metamaterial. 

However, as magnetodielectric MMs can generally be optically anisotropic, 
the reflection matrix ${\bf R}_2$ for the MM-vacuum interface is in general not given by the usual Fresnel formulas (\ref{FresnelCoefficientsMM}). In their most general form metamaterials can be bi-anisotropic, meaning that the constitutive relations have the form \cite{Kong}
\begin{eqnarray}
\label{bianisotropy}
{\bf D} = \boldsymbol{\epsilon} \cdot {\bf E} +  \boldsymbol{\kappa} \cdot {\bf H} , \\
{\bf B} =  \boldsymbol{\zeta} \cdot {\bf E} + \boldsymbol{\mu} \cdot {\bf H}.
\end{eqnarray}     
Here $\boldsymbol{\kappa}$ and $\boldsymbol{\zeta}$ are the magneto-optical permittivities, and they describe magnetic-electric
cross-coupling. There are indeed some metamaterials in which the magneto-optical tensors $\boldsymbol{\kappa}$ and $\boldsymbol{\zeta}$ are not negligible \cite{marques}, but since these properties can be almost entirely suppressed by using a sufficiently symmetric unit cell \cite{Padilla}, we assume henceforth that 
${\bf D} = \boldsymbol{\epsilon} \cdot {\bf E}$ and ${\bf B} = \boldsymbol{\mu} \cdot {\bf H}$. We also assume again 
that the material tensors $\boldsymbol{\epsilon}$ and $\boldsymbol{\mu}$ are functions of frequency only, neglecting any possible spatial dispersion.  

Even without bi-anisotropy the physics of (uni)anisotropic materials is still very rich \cite{Chew, Visnovsky}. It is very common to describe them according to their degree of symmetry; in crystallographic theory this leads to Bravais lattices and their associated point groups \cite{Kittel}. This classification is also very useful for the study of metamaterials, since they may usually be described in terms of unit cells (split-ring-resonators \cite{PendrySRR}, nanopillars \cite{grigorenko}, nanorods \cite{shalaev}, nanospheres \cite{Yannopapas,Wheeler}, etc.) arranged in a periodic lattice. The most extreme anisotropic situation is when the only symmetry of the unit cell is inversion with respect to the origin. In this case, known as the triclinic system, both the permittivity and the permeability tensors have nine non-zero components \cite{Visnovsky} in a given orthogonal coordinate system, making the formulation very cumbersome. Although it is certainly possible to diagonalize at least one of the tensors by choosing a suitable basis, the angles formed by the eigenvectors depend upon frequency in the triclinic system \cite{LandauContMedia, BornWolf}. 
Since the force (\ref{eq1}) is an integral over all frequencies, this frequency-dependent diagonalization is of little help for purposes of calculating Casimir forces. Fortunately, it is still possible to investigate anisotropic effects in the Casimir force without going into such an involved case, so we restrict ourselves in the next two subsections to basically two types of anisotropy.

%%%%%%%%%%%%%%%%%%

\subsection{Reflection matrices for uniaxial (out-of-plane) planar metamaterials}

In this subsection we calculate the reflection matrix ${\bf R}_2$ for the case of a planar interface between vacuum and a uniaxial magnetodielectric medium that is isotropic on the interface plane, {\it i.e.}, whose electric and magnetic anisotropic directions coincide and are perpendicular to the interface. In optical terminology, this is an example of a uniaxial medium \cite{BornWolf} with the optic axis coinciding with the anisotropic direction. It is known that for uniaxial lattices, that is, the ones belonging either to the trigonal, tetragonal and hexagonal crystallographic systems \cite{Kittel}, the electromagnetic tensors are diagonal in the coordinate system defined by any two orthogonal directions in the symmetry plane and the optic axis 
\cite{LandauContMedia}. Therefore, choosing the interface as the $xy$ plane and the anisotropic medium to be the half-space defined by ${\it z}>0$ (see Fig. 2), the permittivity and permeability tensors are given by 
\begin{eqnarray}
\label{anisotropy}
\epsilon_{ij} = \left[ 
\begin{array}{ccc}
\epsilon_{xx} & 0 & 0 \\
0 & \epsilon_{xx}  & 0 \\
0 & 0 & \epsilon_{zz}
\end{array} \right] 
\;\; ; \;\;
\mu_{ij} = \left[ 
\begin{array}{ccc}
\mu_{xx} & 0 & 0 \\
0 & \mu_{xx}  & 0 \\
0 & 0 & \mu_{zz}
\end{array} \right] ,
\end{eqnarray}
where we have used $\epsilon_{yy}=\epsilon_{xx}$ and $\mu_{yy}=\mu_{xx}$, whose frequency dependence is
implicit. Although the calculation of the reflection matrix for a metamaterial
with a single out-of-plane anisotropic direction is relatively simple and akin to the isotropic case,
for the sake of completeness we briefly review this calculation, which is relevant to several metamaterials having such anisotropy \cite{Zhang}. 

%
%
%%%%%%%%%%%%%%%%%%%%%%%%%%%%%%%%%%%%%%%%%%%%%
%%%%%%%%%%%%%%%%%   DESENHO ONDA INCIDENTE  UNIAXIAL %%%%%%%%%%%%%%%%%%%%%%%%%%%%%%%%%%%%%%%%%%%%%%%%%%%%%%%%%%%%%%%%%
%
\begin{figure}
\begin{center}
\scalebox{0.5}{\includegraphics{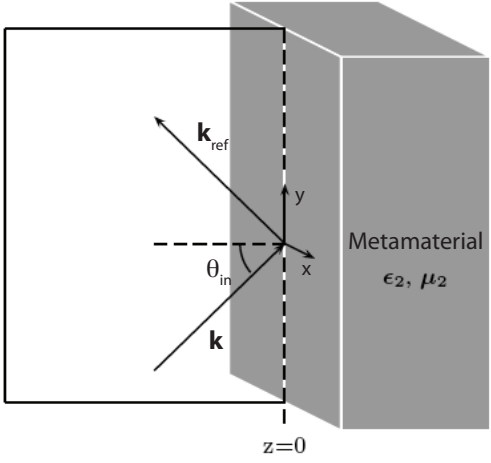}}
\caption{An incident plane wave impinging on a uniaxial metamaterial with its optic axis perpendicular to the $z=0$ plane.}
\label{fig2}
\end{center}
\end{figure}
%

%%%%%%%%%%%%%%%%%%%%%%%%%%%%%%%%%%%%%%%
%%%%%%%%%%%%%%%%%%%%%%%%%%%%%%%%%%%%%%%%
%%%%%%%%%%%%%%%%%%%%%%%%%%%%%%%%%%%%%

Let us assume that a plane wave with wave vector ${\bf k}$ and polarization $p$ impinges upon the interface from \linebreak ${z}<0$ (region 3, vacuum) towards the metamaterial (region 2). Given the rotational symmetry about $z$, without loss of generality we can choose our coordinate system so that the plane of incidence (defined by ${\bf k}$ and ${\bf z}$) coincides with the $xz$ plane (see Fig. 2). In order to solve the reflection-refraction problem we have to know how waves propagate in the anisotropic medium. In this particular case of uniaxial anisotropy orthogonal to the interface it may be shown by direct substitution that
TE waves
\begin{equation}
{\bf E}^{\rm TE} = E_0 {\bf y} e^{i (k_x x + k_z z) } e^{-i \omega t},
\end{equation}
are solutions to Maxwell's equations provided that
\begin{eqnarray}
\frac{k_{2,x}^2}{\mu_{zz}} + \frac{k_{2,z}^2}{\mu_{xx}}= \frac{\omega^2}{c^2} \epsilon_{xx} \;
[\mbox{in the MM}]  \\
k_{1,x}^2 + k_{1,z}^2 = \frac{\omega^2}{c^2}  \;
[\mbox{in vacuum}] .
\end{eqnarray}
In a similar fashion, TM waves 
\begin{equation}
{\bf H}^{\rm TM} = H_0 {\bf y} e^{i (k_x x + k_z z) } e^{-i \omega t} ,
\end{equation}
are solutions to Maxwell's equations provided that
\begin{eqnarray}
\frac{k_{2,x}^2}{\epsilon_{zz}} + \frac{k_{2,z}^2}{\epsilon_{xx}}= \frac{\omega^2}{c^2} \mu_{xx} \;
[\mbox{in the MM}]  \\
k_{1,x}^2 + k_{1,z}^2 = \frac{\omega^2}{c^2}  \;
[\mbox{in vacuum}] .
\end{eqnarray}
Therefore there is no polarization-mixing, and consequently the off-diagonal elements of the reflection matrix vanish:
$r_2^{\rm TE,TM}(i \xi, {\bf k}_{\|})=r_2^{\rm TM,TE}(i \xi, {\bf k}_{\|})=0$.
This allows one to consider separately the reflection of TE and TM waves. 

Let  ${\bf E}_{{\rm in}} = E_0 {\bf y} e^{i (k_{1,x} x + k_{1,z} z)}$ be a TE field incident from the vacuum side. Given the translational invariance of the material
properties along the planar interface, only specular 
reflection occurs, which implies that both $x$ and $y$ components of the wave vector ${\bf k}$ are continuous.
Therefore, the reflected TE field is ${\bf E}_{{\rm ref}} = r_2 E_0 {\bf y} e^{i (k_{1,x} x - k_{1,z} z)}$, and the
transmitted TE field is
${\bf E}_{{\rm t}} = t_2 E_0 {\bf y} e^{i (k_{1,x} x + k_{2,z} z)}$, with 
$k_{2,z}^2 = (\omega^2/c^2) \epsilon_{xx} \mu_{xx} - k_{1,x}^2 \mu_{xx} / \mu_{zz}$. 
Imposing the boundary conditions on the TE modes, we have
\begin{eqnarray}
E_{{\rm in},y} + E_{{\rm ref},y} &=& E_{{\rm t},y}
\; \Rightarrow \; 1+r_2 = t_2 ,  \nonumber  \\
H_{{\rm in},x} + H_{{\rm ref},x} = H_{{\rm t},x} \hspace{-3pt} & \Rightarrow& \hspace{-3pt}
(-1+r_2 )k_{1,z}  = -t_2 \frac{k_{2,z}}{\mu_{xx}} ,
\nonumber 
\end{eqnarray} 
from which it follows that
$r_2 = (\mu_{xx} k_{1,z} - k_{2,z})/(\mu_{xx} k_{1,z} + k_{2,z})$.
Evaluating this expression along imaginary
frequencies $\omega= i \xi$, one obtains the TE-TE reflection
amplitude on the vacuum-MM interface \cite{uniaxialRC}:
\begin{eqnarray}
\label{AnisotropicFresnelCoefficients1}
r_{2_{\rm uni}}^{\rm TE,TE}(i \xi, {\bf k}_{\|}) = \frac{\mu_{xx} K_3 - \sqrt{\frac{\mu_{xx}}{\mu_{zz}} k_{\|}^2 + \mu_{xx} \epsilon_{xx} 
\frac{\xi^2}{c^2} }}{\mu_{xx} K_3 + \sqrt{\frac{\mu_{xx}}{\mu_{zz}} k_{\|}^2 + \mu_{xx} \epsilon_{xx}
\frac{\xi^2}{c^2} }}, 
\end{eqnarray}
where, we recall,  $K_3=\sqrt{k_{\|}^2 + \xi^2/c^2}$ and $k_{\|}^2=k_x^2+k_y^2$. 
Following similar steps, the TM reflection amplitude on the vacuum-MM interface can also be derived:
\begin{eqnarray}
\label{AnisotropicFresnelCoefficients2}
r_{2_{\rm uni}}^{\rm TM,TM}(i \xi, {\bf k}_{\|}) = \frac{\epsilon_{xx} K_3 - \sqrt{\frac{\epsilon_{xx}}{\epsilon_{zz}} k_{\|}^2 + \epsilon_{xx} \mu_{xx} 
\frac{\xi^2}{c^2} }}{\epsilon_{xx} K_3 + \sqrt{\frac{\epsilon_{xx}}{\epsilon_{zz}} k_{\|}^2 + \epsilon_{xx} \mu_{xx}
\frac{\xi^2}{c^2} }}.
\end{eqnarray}

%%%%%%%%%%%%%%%%%%%%%%%%%%%%%%%%%%%%%%%%

\subsection{Reflection matrices for biaxial, anisotropic magnetodielectrics}

In ascending order of symmetry, the crystals belonging to the triclinic, monoclinic and orthorhombic crystallographic systems \cite{Kittel} are known as biaxial crystals, since they are characterized by two optic axes. In this subsection we shall restrict ourselves to the orthorhombic case \cite{TeltlerHenvis}, which allows simultaneous diagonalization of $\boldsymbol{\epsilon}$ and $\boldsymbol{\mu}$ in an orthonormal basis. The calculation of the reflection matrices for the other two types of biaxial metamaterials is conceptually equivalent but more cumbersome since the material tensors cannot be brought to diagonal form in a frequency-independent basis. 

Let us then consider the system described in Fig. 3, which is similar to Fig. 2 but with an orthorhombic metamaterial on the right side. Assuming it is possible to prepare the MM in such a way that one of the eigenvectors is perpendicular to the interface, then the diagonal basis is just $\{ \hat{\bf x}, \hat{\bf y}, \hat{\bf z} \}$ and the electromagnetic tensors are given by     
\begin{eqnarray}
\label{OrthorhombicTensors}
\epsilon_{ij} = \left[ 
\begin{array}{ccc}
\epsilon_{xx} & 0 & 0 \\
0 & \epsilon_{yy}  & 0 \\
0 & 0 & \epsilon_{zz}
\end{array} \right] 
\;\; ; \;\;
\mu_{ij} = \left[ 
\begin{array}{ccc}
\mu_{xx} & 0 & 0 \\
0 & \mu_{yy}  & 0 \\
0 & 0 & \mu_{zz}
\end{array} \right] .
\end{eqnarray}
Several metamaterials can be described by material tensors like (\ref{OrthorhombicTensors}); a good example is the fishnet design used in \cite{dolling}.
%
%
%%%%%%%%%%%%%%%%%%%%%%%%%%%%%%%%%%%%%%%%%%%%%
%%%%%%%%%%%%%%%%%   DESENHO ONDA INCIDENTE   BIAXIAL%%%%%%%%%%%%%%%%%%%%%%%%%%%%%%%%%%%%%%%%%%%%%%%%%%%%%%%%%%%%%%%%%
%
\begin{figure}
\begin{center}
\scalebox{1}{\includegraphics{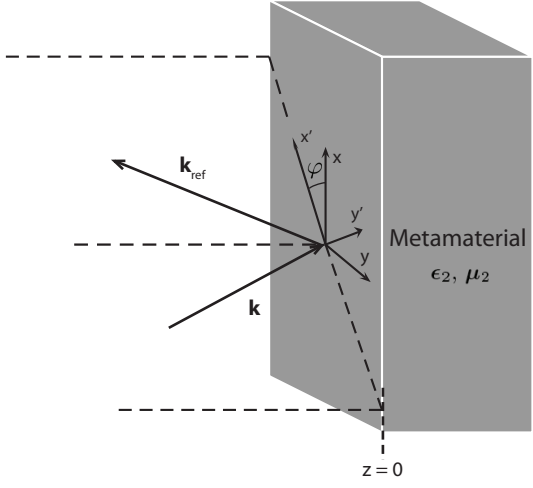}}
\caption{An incident plane wave impinging on a biaxial metamaterial with orthorhombic symmetry (see text).}
\label{fig3a}
\end{center}
\end{figure}
%

%%%%%%%%%%%%%%%%%%%%%%%%%%%%%%%%%%%%%%%

Metamaterials with two optic axes, even those with the simplest orthorhombic symmetry, are much harder to treat than those with an out-of-plane, uniaxial optic axis described in the previous subsection. The reason is that Maxwell's equations do not support transverse waves for biaxial materials: neither TE nor TM
waves are solutions inside the material, and the off-diagonal elements of the reflection matrix do not vanish. 
This also happens for uniaxial materials with in-plane optic axes, whose reflection matrix can be obtained as a particular case of orthorhombic materials with $\epsilon_{yy}=\epsilon_{zz}$ and $\mu_{yy}=\mu_{zz}$. The Casimir interaction between two dielectric semi-spaces with one in-plane optic axis was treated in \cite{BarashAnisotropy} and used in the experimental proposal to measure the Casimir torque between birefrigent plates \cite{Casimirtorque}.

The calculation of the plane-wave solutions to Maxwell's
equations is simplified using a coordinate system attached to an incident wave
from the vacuum side. Let a plane wave with incident wave vector ${\bf k}$ impinge on the interface forming an angle $\theta_{\rm in}$ with the normal direction (see Fig. 3). Let $(x',y',z')$ be the coordinate system attached to the corresponding plane of incidence, that forms an angle 
$\varphi$ with the $x$ axis. The optical tensors in this new coordinate system are
\begin{widetext}
\begin{equation}
\epsilon_{i'j'} = \left[ 
\begin{array}{ccc}
\epsilon_{xx} \cos^2\varphi  +  \epsilon_{yy} \sin^2 \varphi & 
(\epsilon_{xx}-\epsilon_{yy}) \sin\varphi \cos\varphi & 
0 \\
(\epsilon_{xx}-\epsilon_{yy}) \sin\varphi \cos\varphi & 
\epsilon_{xx} \sin^2\varphi  +  \epsilon_{yy} \cos^2 \varphi & 
0 \\
0 &
0 & 
\epsilon_{zz}
\end{array} \right] ;
\nonumber
\end{equation}
and
\begin{equation}
\mu_{i'j'} = \left[ 
\begin{array}{ccc}
\mu_{xx} \cos^2\varphi  +  \mu_{yy} \sin^2 \varphi & 
(\mu_{xx}-\mu_{yy}) \sin\varphi \cos\varphi & 
0 \\
(\mu_{xx}-\mu_{yy}) \sin\varphi \cos\varphi & 
\mu_{xx} \sin^2\varphi  +  \mu_{yy} \cos^2 \varphi & 
0 \\
0 &
0 & 
\mu_{zz}
\end{array} \right] .
\nonumber
\end{equation}
\end{widetext}

The expressions for the incident fields are
\begin{equation}
\label{incidentelectric}
{\bf E}_{\rm in} = \left[e^{\rm TE}_{\rm in} \hat{\bf y}' + 
e^{\rm TM }_{\rm in} \frac{c}{\omega} (q_{\rm in}\hat{\bf x}' - k_{x'} \hat{\bf z}')\right]  e^{i (k_{x'} x' + q_{\rm in} z' - \omega t)} ,
\end{equation} 
\begin{equation}
\label{incidentmagnetic}
{\bf H}_{\rm in} = \left[e^{\rm TM}_{\rm in} \hat{\bf y}' - 
e^{\rm TE }_{in} \frac{c}{\omega} (q_{\rm in}\hat{\bf x}' - k_{x'} \hat{\bf z}')\right] e^{i (k_{x'} x' + q_{\rm in} z' - \omega t)} , 
\end{equation} 
where $e^{\rm TE}_{\rm in}$, $e^{\rm TM}_{\rm in}$ are given amplitudes and we defined 
$k_{x'}= (\omega/c) \sin \theta_{\rm in}$ and $q_{\rm in}= (\omega/c) \cos \theta_{\rm in}$. 
The reflected wave has a similar expression:
\begin{equation}
\label{reflectedelectric}
{\bf E}_{\rm ref} = \left[e^{\rm TE}_{\rm ref} \hat{\bf y}' - 
e^{\rm TM }_{\rm ref} \frac{c}{\omega} (q_{\rm in}\hat{\bf x}' + 
k_{x'} \hat{\bf z}')\right] e^{i (k_{x'} x' - q_{\rm in} z' - \omega t)} ,
\end{equation} 
\begin{equation}
\label{reflectedmagnetic}
{\bf H}_{\rm ref} = \left[e^{\rm TM}_{\rm ref} \hat{\bf y}' + e^{\rm TE}_{\rm ref} \frac{c}{\omega} (q_{\rm in}\hat{\bf x}' + k_{x'} \hat{\bf z}')\right] e^{i (k_{x'} x' - q_{\rm in} z' - \omega t)} , 
\end{equation}
where we have used $q_{\rm ref} = -q_{\rm in}$. Our problem now consists in finding the amplitudes $e^{\rm TE}_{\rm ref}$ $e^{\rm TM}_{\rm ref}$, so we can construct the reflection matrix (\ref{ReflectionMatrices}). In order to obtain the reflection amplitudes, however, it is necessary find the transmitted fields as well, which means that we have to solve Maxwell's equations in the metamaterial. 

Let us assume plane waves 
\begin{eqnarray}
{\bf E} = {\bf e}(z') e^{i ( k_{x'} x' - \omega t)}  &;& {\bf e}=(e_{x'},e_{y'},e_{z'}) , \nonumber \\  
{\bf H} = {\bf h}(z') e^{i ( k_{x'} x' - \omega t)} &;& {\bf h}=(h_{x'},h_{y'},h_{z'}) , 
\label{planewaves}
\end{eqnarray}
as solutions to Maxwell's equations in medium 2, where we have already deduced the $x'$ dependence from the phase-matching condition on the interface ($k_{x'}$ is conserved across the interface). 
By substituting (\ref{planewaves}) into the Faraday and Amp\`ere-Maxwell laws
\begin{eqnarray}
\label{FaradayAMLaws}
\nabla \times {\bf E} = -\frac{1}{c} \frac{\partial {\bf B}}{\partial t}\;\; , \;\; 
\nabla \times ({\boldsymbol{\mu}^{-1} \cdot {\bf B}}) = \frac{1}{c} \frac{\partial ({\boldsymbol{\epsilon} \cdot {\bf E}})}{\partial t} ,
\end{eqnarray}
respectively, we see that the $z'$ components can be eliminated as 
\begin{eqnarray}
e_{z'}=- c k_{x'} h_{y'}/ \omega \epsilon_{z'z'} \;\; ; \;\; 
h_{z'}= c k_{x'} e_{y'}/ \omega \mu_{z'z'} \ .
\end{eqnarray}
In order to determine the remaining $x'$ and $y'$ components of ${\bf e}$ and ${\bf h}$ it is convenient to introduce a vector ${\bf u}$ with components $u_1=e_{x'}$, $u_2=e_{y'}$, $u_3=h_{x'}$ and $u_4=h_{y'}$. 
With the ansatz $u_j=u_j(0) e^{i q z'}$ we obtain the following linear system of equations:
\begin{equation}
\label{MatrixEquation}
{\bf L} \cdot {\bf u} = - \frac{c}{\omega} q \;{\bf u},
\end{equation}
where the non-zero elements of the matrix ${\bf L}$ are:
\begin{eqnarray}
\label{matrixLelements}
L_{13}=-L_{24}=  - (\mu_{xx} - \mu_{yy}) \sin\varphi \cos\varphi , \nonumber \\
L_{14} =  \frac{k_{x'}^2 c^2}{ \omega^2 \epsilon_{zz}} - \mu_{yy} \cos^2\varphi - 
\mu_{xx} \sin^2\varphi , \nonumber \\
L_{23}= \mu_{xx} \cos^2\varphi +  \mu_{yy} \sin^2\varphi , \nonumber \\
L_{31}=-L_{42}= (\epsilon_{xx} - \epsilon_{yy}) \sin\varphi \cos\varphi , \nonumber \\
L_{32}=- \frac{k_{x'}^2 c^2}{\omega^2 \mu_{zz}} + \epsilon_{yy} \cos^2\varphi + \epsilon_{xx} \sin^2\varphi ,\nonumber \\
L_{41}= -\epsilon_{xx} \cos^2\varphi - \epsilon_{yy} \sin^2\varphi .
\end{eqnarray}
The condition for non-trivial solutions ($ {\rm det} ({\bf L}+ \omega q/c) = 0$)
gives us the equation that determines the possible values of $q$, namely
\begin{equation}
\label{bi-quadratic}
\left( \frac{c^2}{\omega^2}q^2-A \right) \left( \frac{c^2}{\omega^2}q^2-B \right) = C, 
\end{equation}
where
\begin{eqnarray}
A &=& L_{13} L_{31} + L_{14} L_{41} , \nonumber \\
B &=& L_{23} L_{32} + L_{24} L_{42} , \nonumber \\
C_1 &=& L_{13} L_{32} + L_{14} L_{42} , \nonumber \\
C_2 &=& L_{23} L_{31} + L_{24} L_{41} , \nonumber \\
C &=& C_1 C_2 ,
\nonumber
\end{eqnarray}
whose four solutions $q^{(m)}$ ($m=1,2,3,4$) are
\begin{equation}
\label{bi-solutions}
q^{(m)} = \pm \frac{\omega}{c} \frac{1}{\sqrt{2}} \sqrt{ A+B \pm \sqrt{(A-B)^2 + 4 C}} .
\end{equation}
These solutions may be conveniently split into two pairs, according to the sign of ${\rm Re}$ $q^{(m)}$ - solutions with ${\rm Re}$ $q^{(m)} > 0$ (${\rm Re}$ $q^{(m)} < 0$) define positive (negative) propagating waves. If we denote the positive solutions by $m=1,2$, we may write the general solution for ${\bf u}$ as
\begin{eqnarray}
{\bf u}(z') &=& \sum_{m=1,2} {\bf u}^{(m)}(0) \; e^{i q^{(m)} z'} \nonumber \\
&+& \sum_{m=3,4} {\bf u}^{(m)}(0) \; e^{-i q^{(m-2)} z'} ,
\end{eqnarray}
where we have used $q^{(3)} =  -q^{(1)}$ and $ q^{(4)} =  -q^{(2)}$. It is easy to see that the refraction of a wave coming from $z'<0$ can only give rise to positive propagating waves (in the sense defined above), from which we conclude that ${\bf u}^{(3)}(0) = {\bf u}^{(4)}(0) = 0$. Therefore, the transmitted field into the anisotropic magnetodielectric medium is
\begin{eqnarray}
\label{transmittedfields}
\left[ \begin{array}{c}
{\bf E}_{\rm t} \\
{\bf H}_{\rm t}
\end{array} \right]
= e^{i (k_{x'} x' - \omega t)}  \!\!\! \sum_{m=1,2} {\bf u}^{(m)}(0) \; e^{i q^{(m)} z'} .
\end{eqnarray}

In order to find the amplitudes ${\bf u}^{(m)}(0)$ we have to impose the proper boundary conditions on the fields. In this case they just require the continuity of $E_{x'}, E_{y'}, H_{x'}, H_{y'}$ across the interface.
Using (\ref{incidentelectric})-(\ref{reflectedmagnetic}) and (\ref{transmittedfields}), one derives the following
boundary conditions:
\begin{eqnarray}
\label{system}
q_{\rm in} (e^{\rm TM }_{\rm in} - e^{\rm TM }_{\rm ref}) &=& \frac{\omega}{c} \sum_{m=1,2}  e^{(m)}_{x'}(0) , \nonumber \\ 
e^{\rm TE }_{\rm in} + e^{\rm TE }_{\rm ref} &=& \sum_{m=1,2}  e^{(m)}_{y'}(0) , \nonumber \\
- q_{\rm in} (e^{\rm TE}_{\rm in} - e^{\rm TE}_{\rm ref}) &=& \frac{\omega}{c} \sum_{m=1,2}  h^{(m)}_{x'}(0) , \nonumber \\
e^{\rm TM}_{\rm in} + e^{\rm TM}_{\rm ref} &=& \sum_{m=1,2} h^{(m)}_{y'}(0) .
\end{eqnarray}
This system of equations is unsolvable as it stands, given the large number of unknowns. It is possible, however, to 
use (\ref{MatrixEquation}) to express all the transmitted amplitudes in terms of just one, say
$e^{(m)}_{x'}$:
\begin{eqnarray}
&&\alpha^{(m)} \equiv  \frac{e^{(m)}_{y'}(0)}{e^{(m)}_{x'}(0)} = \frac{(q^{(m)})^2 - (\omega^2/c^2)A}{(\omega^2/c^2)C_1} , \nonumber \\
&&\beta^{(m)} \equiv \frac{h^{(m)}_{x'}(0)}{e^{(m)}_{x'}(0)} = - \frac{\omega}{c}\frac{L_{31}}{q^{(m)}} - \frac{\omega}{c}\frac{L_{32}}{q^{(m)}} \alpha^{(m)} ,\nonumber \\
&&\gamma^{(m)} \equiv  \frac{h^{(m)}_{y'}(0)}{e^{(m)}_{x'}(0)} = -\frac{\omega}{c}\frac{L_{41}}{q^{(m)}} - \frac{\omega}{c}\frac{L_{42}}{q^{(m)}} \alpha^{(m)}. \nonumber 
\end{eqnarray} 
Using these definitions, we can rewrite (\ref{system}) as
\begin{widetext}
\begin{eqnarray}
\label{system2}
\left[ \begin{array}{cccc}
  -1 & 0 & \alpha^{(1)} & \alpha^{(2)} \\    
  c q_{\rm in}/\omega & 0 & -\beta^{(1)} & -\beta^{(2)} \\
  0 & c q_{\rm in}/\omega & 1 & 1 \\
  0 & -1 & \gamma^{(1)} & \gamma^{(2)}  
\end{array} \right]
\left[ \begin{array}{c}
   e^{\rm TE }_{\rm ref}      \\
   e^{\rm TM }_{\rm ref}    \\
   e^{(1)}_{x'}(0) \\
   e^{(2)}_{x'}(0)   
\end{array} \right] 
=
\left[ \begin{array}{c}
     e^{\rm TE }_{\rm in}      \\
   c q_{\rm in}/\omega e^{\rm TE }_{\rm in}    \\
   c q_{\rm in}/\omega e^{\rm TM }_{\rm in}  \\
    e^{\rm TM }_{\rm in}    
\end{array} \right] .
\end{eqnarray} 
\end{widetext}

In order to find the reflection coefficients, we must solve (\ref{system2}) for the reflected amplitudes. For the sake of clarity, let us do this separately for $ e^{\rm TM }_{\rm in} = 0, \ e^{\rm TE }_{\rm in} \neq 0$ and for 
$e^{\rm TM }_{\rm in} \neq 0 , \ e^{\rm TE }_{\rm in} = 0$. In the first case, Cramer's rule immediately yields
\begin{eqnarray}
\label{solutionTE1}
r_2^{\rm TE,TE}(i \xi, {\bf k}_{\|}) &=& \frac{e^{\rm TE}_{\rm ref}}{e^{\rm TE}_{\rm in}}=
\left. \frac{\det \mathbb{M}_1}{\det \mathbb{M}}  \right|_{\stackrel{\omega \rightarrow i \xi}{k_{x'}\rightarrow k_{\|}}}, \\
\label{solutionTE2}
r_2^{\rm TM,TE} (i \xi, {\bf k}_{\|}) &=& \frac{e^{\rm TM}_{\rm ref}}{e^{\rm TE}_{\rm in}}= \left. \frac{\det \mathbb{M}_2}{\det \mathbb{M}}  \right|_{\stackrel{\omega \rightarrow i \xi}{k_{x'}\rightarrow k_{\|}}} ,
\end{eqnarray}
and in the second case we have
\begin{eqnarray}
\label{solutionTM1}
r_2^{\rm TE,TM} (i \xi, {\bf k}_{\|}) &=& \frac{e^{\rm TE}_{\rm ref}}{e^{\rm TM}_{\rm in}}= \left. \frac{\det \mathbb{M}_3}{\det \mathbb{M}}  \right|_{\stackrel{\omega \rightarrow i \xi}{k_{x'}\rightarrow k_{\|}}}, \\
\label{solutionTM2}
r_2^{\rm TM,TM} (i \xi, {\bf k}_{\|}) &=& \frac{e^{\rm TM}_{\rm ref}}{e^{\rm TM}_{\rm in}}= \left. \frac{\det \mathbb{M}_4}{\det \mathbb{M}}  \right|_{\stackrel{\omega \rightarrow i \xi}{k_{x'}\rightarrow k_{\|}}},
\end{eqnarray}
where $\mathbb{M}$ is the $4\times 4$ matrix in (\ref{system2}) and
\begin{eqnarray}
\label{matricesM}
&&\mathbb{M}_1 = \left[ \begin{array}{cccc}
  1 & 0 & \alpha^{(1)} & \alpha^{(2)} \\    
  c q_{\rm in}/\omega & 0 & -\beta^{(1)} & -\beta^{(2)} \\
  0 & c q_{\rm in}/\omega  & 1 & 1 \\
  0 & -1 & \gamma^{(1)} & \gamma^{(2)}  
\end{array} \right] , \nonumber \\
&&\mathbb{M}_2 = \left[ \begin{array}{cccc}
  -1 & 1 & \alpha^{(1)} & \alpha^{(2)} \\    
  c q_{\rm in}/\omega  & c q_{\rm in}/\omega  & -\beta^{(1)} & -\beta^{(2)} \\
  0 & 0 & 1 & 1 \\
  0 & 0 & \gamma^{(1)} & \gamma^{(2)}  
\end{array} \right] , \nonumber \\
&&\mathbb{M}_3 = \left[ \begin{array}{cccc}
  0 & 0 & \alpha^{(1)} & \alpha^{(2)} \\    
  0 & 0 & -\beta^{(1)} & -\beta^{(2)} \\
  c q_{\rm in}/\omega  & c q_{\rm in}/\omega  & 1 & 1 \\
  1 & -1 & \gamma^{(1)} & \gamma^{(2)}  
\end{array} \right] ,\nonumber \\
&&\mathbb{M}_4 = \left[ \begin{array}{cccc}
  -1 & 0 & \alpha^{(1)} & \alpha^{(2)} \\    
  c q_{\rm in}/\omega  & 0 & -\beta^{(1)} & -\beta^{(2)} \\
  0 & c q_{\rm in}/\omega  & 1 & 1 \\
  0 & 1 & \gamma^{(1)} & \gamma^{(2)}  
\end{array} \right] .
\end{eqnarray}
 
%%%%%%%%%%%%%%%%%%%%%%%%%%%%%%%%%%%%%%

\section{Metallic-based metamaterials and the Casimir effect} 

Metamaterials may be roughly divided into two classes. The first class consists of MMs that are partially or totally based on metallic structures.  In this section we concentrate on these metallic-based MMs, which were previously considered by us in \cite{usPRL}. We study in detail the effects of optical anisotropy on the Casimir-Lifshitz interaction with magnetodielectric media. The second class consists of MMs based purely on dielectric materials, that we shall treat in the next section.

%%%%%%%%%%%%

\subsection{Isotropic metamaterials} 

Before going straight to the calculations, it is necessary to point out that metallic MMs may be also divided into two types, which we shall characterize as (i) connected and (ii) non-connected. As the name suggests, in the connected MMs the metallic part is partially or totally interconnected throughout the metamaterial \cite{dolling}, while in the non-connected it is not \cite{grigorenko,shalaev}. This distinction is important because in connected MMs there is a net conductivity contribution to the dielectric function due to the metallic part, while in the non-connected MMs the background is effectively non-conducting.

Let us begin with the simple example of a metallic half-space 1
in front of an isotropic, connected metallic-based metamaterial 2. For the metal we assume the usual Drude model
\begin{eqnarray}
\label{Drude}
\epsilon_1(\omega) = 1 - \frac{\Omega_1^2}{(\omega^2 + i \gamma_1 \omega)} \; , \; \mu_1(\omega) = 1 ,
\end{eqnarray}
where $\Omega_1$ is its plasma frequency and $\gamma_1$ the dissipation coefficient. For the second half-space we have to be more specific about the MM we want to consider. In the simplest description isotropic, connected metallic metamaterials may be described by a dielectric response accounting for both a resonance and a Drude contribution: 
\begin{eqnarray}
\label{Drude Background}
\epsilon_2(\omega)  \hspace{-3pt} = 1- (1-f) \frac{\Omega_{e}^2}{\omega^2 - \omega_{e}^2 + i \gamma_{e} \omega}  - f \frac{\Omega_{D}^2}{\omega^2 + i \gamma_{D}} ,
\end{eqnarray}
where $\Omega_{e}$, $\omega_{e}$ and $\gamma_e$ are respectively the effective electric oscillating strength, the resonance frequency, and the effective dissipation parameter of the resonant part, and $\Omega_D$ and $\gamma_D$ are the Drude parameters of the metallic background of the MM.
The filling factor $f$ roughly quantifies the fraction of metallic structure present in the MM. The magnetic permeability is given by a resonant part alone:
\begin{equation}
\label{Drude-Lorentz} 
\mu_2(\omega) = 1 - \frac{\Omega_{m}^2}{ \omega^2 - 
\omega_{m}^2 + i \gamma_{m} \omega} ,
\end{equation}
where $\Omega_{m}$, $\omega_{m}$ and $\gamma_m$ are defined analogously to their electric counterparts. In Fig. 4 we plot the Casimir-Lifshitz force between a metallic half-space and an isotropic metallic-based planar metamaterial described by (\ref{Drude Background}) and (\ref{Drude-Lorentz}) for different filling factors at zero temperature. We see that without the Drude contribution ($f = 0$) there is repulsion for a certain range of distances, as long as the half-space 2 is mainly magnetic ($\epsilon_2(i \xi) < \mu_2(i \xi)$). However, as we {\lq\lq turn on\rq\rq} a metallic background ($f > 0$), the permittivity grows stronger and reverts the previous relation for a larger and larger range of frequencies, up to the point where the magnetic activity is no longer able to produce repulsion. 

An idealization carried throughout the paper is that both the metal and the metamaterial are infinitely long in the $z$-direction. When we have slabs of finite thickness instead of half-spaces, the reflection coefficients change to \cite{henkel}
\begin{eqnarray}
r_{j_{\rm slab}}^{p,p}(i\xi) = r_j^{pp}(i\xi) \frac{1- e^{-2 K_j d_j}}{1 - {r_j^{p,p}}^2(i\xi) e^{-2K_j d_j}} ,
\end{eqnarray}
where $K_j = \sqrt{k_{\|}^2 + \mu_j(i\xi)\epsilon_j(i\xi) \xi^2/c^2}$, $d_j$ is the thickness of the $j$-th slab, and we are assuming that both slabs are surrounded by vacuum. From the previous expression we see that corrections to the half-space reflection coefficients (at imaginary frequencies) are exponentially small when both products $K_1 d_1$ and $K_2 d_2$ are sufficiently large. This basically tells us that estimates on lower bounds for $d_1$ and $d_2$ are actually model dependent (given that $K_j$ depends on the properties of medium $j$), so in order to discuss those estimates we have to be more specific. For a metal described by (\ref{Drude}) and a wave arriving at normal incidence, we have
\begin{eqnarray}
\label{skindepth1}
K_1 d_1 \gg 1 \Rightarrow d_1 \gg \frac{\lambda_p}{2\pi} \frac{\lambda}{\sqrt{\lambda_p^2 + \lambda^2\lambda_d/(\lambda_d+\lambda)}} ,
\end{eqnarray}
where $\lambda_p = 2 \pi c / \Omega_1$, $\lambda_d = 2 \pi c / \gamma_1$ and $\lambda = 2\pi c/\xi$. For high frequencies, we have $\lambda \ll \lambda_d$ and then (\ref{skindepth1}) becomes
\begin{eqnarray}
d_1 \gg \frac{\lambda_p}{2\pi} \frac{\lambda}{\sqrt{\lambda_p^2 + \lambda^2}} \Rightarrow d \gg \frac{\lambda_p}{2\pi} .
\end{eqnarray}
For typical metals $\lambda_p / 2\pi$ is around $10-20 {\rm nm}$, so, at least for high frequencies, the contribution to the Casimir force of a slab some tens of nanometers wide approaches the contribution of a half-space. However, in the opposite limit ($\lambda \gg \lambda_d$), we have
\begin{eqnarray}
d_1 \gg  \frac{\lambda_p}{2\pi} \sqrt{\frac{\lambda}{\lambda_d}} = \frac{c}{\sqrt{4\pi\sigma_0 \xi}} = \sqrt{2} \delta(\xi)
\end{eqnarray}
where $\sigma_0 = \Omega_1^2/\gamma$ is the static conductivity and
$\delta(\xi)$ is the skin depth of the metal at imaginary frequencies. Thus, for long
wavelengths we see that $d$ scales with $\sqrt{\lambda}$, leading to the conclusion that the half-space approximation is not good for sufficiently low frequencies. Fortunately, for typical materials this is no source of concern, since the integration range where $\lambda \gg \lambda_d$ holds is very small compared to the effective integration range, allowing us to push the slab approximation up to very small frequencies with almost no effect in the final result. One might
wonder what happens for oblique incidence, but it is easy to see
that the more oblique the incident angle is the better the estimate
for $d_1$ holds, since $K_1$ gets larger and larger (this only means
that reflection gets easier as the incidence angle gets larger, as
physically expected). The effect of finite thickness in the Casimir effect was the object of several papers \cite{slabs}, notably in \cite{Tomas} where a systematic procedure was developed to deal with any given number of arbitrary slabs. The effect of finite thickness was also studied in the specific context of Casimir force and metamaterials \cite{henkel,irina,Tomas2}, where it was found that having a layer of a MM instead of a half-space reduces the intensity of the repulsion force and also the range of distances where it occurs.

\begin{center}
\begin{figure}
\scalebox{0.32}{\includegraphics{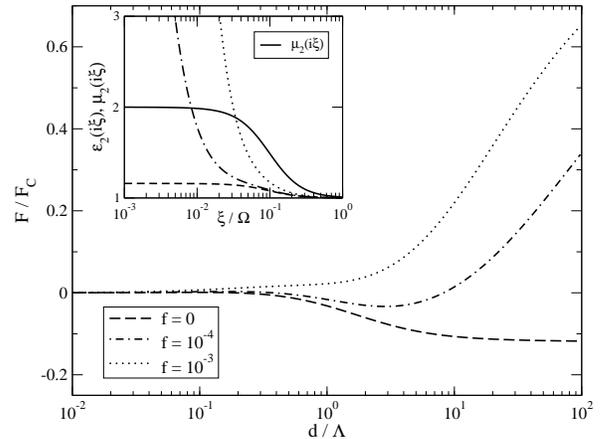}}
\caption{The ratio $F / F_{\rm C}$ for a gold half-space facing an isotropic, interconnected and silver-based metamaterial. $F/A$ is the Casimir force per unit area in this setup, $F_{\rm C} / A = \hbar c \pi^2 / 240 a^4$ is the Casimir force per unit area between two perfect plane conductors, and $F < 0 (F>0)$ corresponds to a repulsive (attractive) force. The frequency scale $\Omega=2 \pi c / \Lambda$ is chosen as the silver plasma frequency $\Omega_D=1.37 \times 10^{16}$ rad/sec. Parameters are: for the metal, $\Omega_1/\Omega=0.96$, $\gamma_1/\Omega=0.004$, and for the metamaterial, 
$\Omega_D/\Omega=1$, $\gamma_D/\Omega=0.006$, 
$\Omega_e/\Omega=0.04$, $\Omega_m/\Omega=0.1$, 
$\omega_e/\Omega=\omega_m/\Omega=0.1$, 
$\gamma_e/\Omega=\gamma_m/\Omega=0.005$.
The inset shows the magnetic permeability $\mu_{2}(i \xi)$
and the electric permittivity $\epsilon_2(i \xi)$ of the MM for the different filling factors.}
\label{fig4}
\end{figure} 
\end{center}

\subsection{Uniaxial Metamaterials} 

Electric anisotropic effects in the Casimir interaction have been thoroughly studied in the literature \cite{BarashAnisotropy, Casimirtorque, ParsegianWeiss, KennethNussinov, Bruno}, but until recently there was no compelling reason to study the consequences of magnetic anisotropy. This changed with the advent of metamaterials, and an investigation of magnetic anisotropy is now in order. The best place to start is to consider uniaxial out-of plane metamaterials, since they constitute the simplest departure from the isotropic case. This type of anisotropy is quite common since it arises naturally when a material is built as a stack of different layers, as is the case for several kinds of MMs \cite{dolling, Zhang, Tanaka, Schilling}. We are particularly interested in the case where the resulting medium is characterized by different degrees of conductivity in the plane of symmetry and in the perpendicular direction to it. 

Let us begin by characterizing the electric and magnetic properties of our uniaxial metamaterial: 
\begin{eqnarray}
\label{EpsilonsMusUniaxial}
&&\epsilon_{xx}(\omega) = \epsilon_{yy}(\omega) = 1-(1-f_x)\frac{\Omega_{e,x}^2}{\omega^2 - \omega_{e,x}^2 + i \gamma_{e,x} \omega} \nonumber \\
&&~~~~~~~~~~~~~~~~~~~~~~~~~~ - f_x \frac{\Omega_{D,x}^2}{\omega^2 + i  \gamma_{D,x}\omega} , \nonumber \\
&&\epsilon_{zz}(\omega)  = 1 - (1-f_z)\frac{\Omega_{e,z}^2}{\omega^2 - \omega_{e,z}^2 + i\gamma_{e,z}\omega} \nonumber \\
&& ~~~~~~~~~~~~~ - f_z \frac{\Omega_{D,z}^2}{\omega^2 + i \gamma_{D,z}\omega} , \nonumber \\
&&\mu_{xx}(\omega) = \mu_{yy}(\omega)= 1 - \frac{\Omega_{m,x}^2}{ \omega^2 - 
\omega_{m,x}^2 + i \gamma_{m,x} \omega} , \nonumber \\
&&\mu_{zz}(\omega) = 1 - \frac{\Omega_{m,z}^2}{ \omega^2 - 
\omega_{m,z}^2 + i \gamma_{m,z} \omega} ,
\end{eqnarray}
where the different filling factors $f_x$ and $f_z$ account for the possible anisotropy in the metallic character of the MM. As we have seen in subsection II-B, the reflection matrix for such a metamaterial is diagonal and given  by (\ref{AnisotropicFresnelCoefficients1}) and (\ref{AnisotropicFresnelCoefficients2}).
\begin{figure}[t]
\hspace{-10pt}
\scalebox{0.58}{\includegraphics{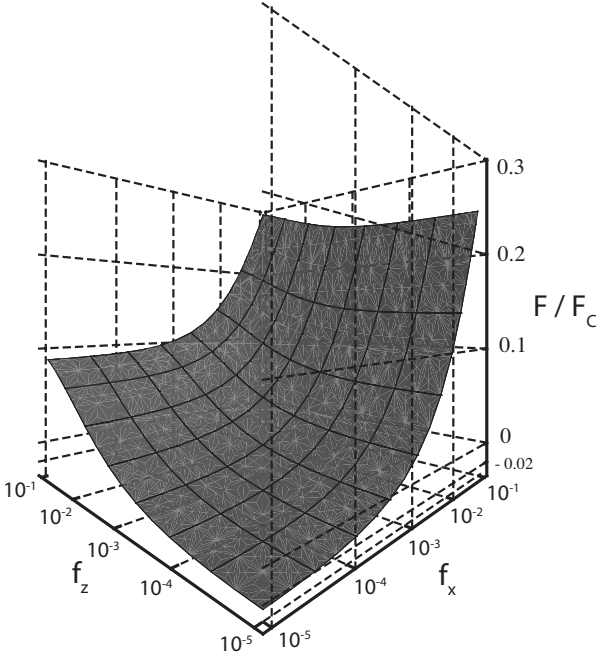}} 
%\scalebox{0.32}{\includegraphics{Fig2b.eps}}  
\caption{The effects of uniaxial anisotropy in the Casimir force between a gold semi-space and a metallic-based connected MM with weak Drude background. The distance is fixed to $d=\Lambda$ and repulsion corresponds to negative values of $F / F_{\rm C}$. All parameters are the same as in Fig. 4 except for the filling factors $f_x$ and $f_z$, which are the variables in this plot.}
\end{figure}
For metallic-based metamaterials with large in-plane electric response $\epsilon_{xx}(i \xi) \gg 1$ at low frequencies, it is clear from Eqs. (\ref{AnisotropicFresnelCoefficients1}, \ref{AnisotropicFresnelCoefficients2})
that anisotropy plays a negligible role in the determination of the reflection coefficients
when there is a dominant Drude background. In order to better appreciate the effects 
of anisotropy we assume henceforth a small or vanishing Drude contribution. In Fig. 5 we show the Casimir force for a metamaterial that has only electric anisotropy ($\mu_{xx}=\mu_{zz}$), which is completely coded in different filling factors ($f_{x} \neq f_{z}$, all other parameters being the same). We see that a repulsive force ($F/F_{\rm C}$) arises only for considerably small values of both $f_{x}$ and $f_{z}$, from which we conclude that killing the Drude background in the $z$-direction alone is not enough to produce Casimir repulsion.

%%%%%%%%%%% 
    
\subsection{Biaxial metamaterials} 

Continuing our track towards more complicated media, we now tackle the biaxial orthorhombic case. Let us consider a metamaterial characterized by the following dielectric and magnetic functions in the basis defined by its eigenvectors (see subsection II-C):
\begin{eqnarray}
\label{EpsilonsMusBiaxial}
&&\epsilon_{xx}(\omega)  = 1 -  (1-f_x)\frac{\Omega_{e,x}^2}{\omega^2 - \omega_{e,x}^2 + i\gamma_{e,x}\omega} \nonumber \\
&& ~~~~~~~~~~~~~~- f_x \frac{\Omega_{D,x}^2}{\omega^2 +  i\gamma_{D,x}\omega} , \nonumber \\
&&\epsilon_{yy}(\omega)  = 1 - (1-f_y)\frac{\Omega_{e,y}^2}{\omega^2 - \omega_{e,y}^2 + i\gamma_{e,y}\omega} \nonumber \\
&& ~~~~~~~~~~~~~~~~ - f_y \frac{\Omega_{D,y}^2}{\omega^2 + i\gamma_{D,y}\omega} , \nonumber \\
&&\epsilon_{zz}(\omega)  = 1 - \frac{\Omega_{e,z}^2}{\omega^2 - \omega_{e,z}^2 + i\gamma_{e,z}\omega} - \frac{\Omega_{D,z}^2}{\omega^2 + i\gamma_{D,z}\omega} , \nonumber \\
&&\mu_{xx}(\omega) = \mu_{yy}(\omega) = 1 - \frac{\Omega_{m,x}^2}{ \omega^2 - 
\omega_{m,x}^2 + i\gamma_{m,x} \omega} , \nonumber \\
&&\mu_{zz}(\omega) = 1 .
\end{eqnarray}
We are particularly interested in the case where $\epsilon_{xx}$ is close to $\epsilon_{yy}$ but in general significantly different from $\epsilon_{zz}$. This means basically that the MM is only slightly anisotropic in the plane of incidence. Our motivation in studying this  particular limiting case is that it is a good approximation for certain types of metamaterials, such as  those based on fishnet designs \cite{dolling}. Note that we are already assuming magnetic in-plane isotropy, which is consistent with a small electric in-plane anisotropy. We may then rewrite the material tensors as
\begin{eqnarray}
\label{OrthorhombicTensors1stOrder}
&&\epsilon_{ij} = \left[ 
\begin{array}{ccc}
\epsilon_{xx} & 0 & 0 \\
0 & \epsilon_{xx}(1 + \delta)  & 0 \\
0 & 0 & \epsilon_{zz}
\end{array} \right] 
\;\; , \nonumber \\
&&\mu_{ij} = \left[ 
\begin{array}{ccc}
\mu_{xx} & 0 & 0 \\
0 & \mu_{xx}  & 0 \\
0 & 0 & 1
\end{array} \right] ,
\end{eqnarray}
where $\delta(\omega) = (\epsilon_{yy}(\omega) - \epsilon_{xx}(\omega))/\epsilon_{xx}(\omega) \ll 1$, and perform the calculations only up to first order in $\delta$. The evaluation of the determinants in (\ref{solutionTE1})-(\ref{solutionTM2}) requires the knowledge of matrix elements of ${\bf L}$ defined by equation (\ref{matrixLelements}) and also of the solutions of equation (\ref{bi-quadratic}). This last step is simplified at first order in $\delta$ because $C/\epsilon_{xx}^2 \sim (\epsilon_{xx}-\epsilon_{yy})^2/\epsilon_{xx}^2 \sim \delta^2$. Therefore
\begin{eqnarray}
\label{Solutions1stOrder}
q^{(1)} = \frac{\omega}{c}\sqrt{A} \;\;\; ; \;\;\; q^{(2)} = \frac{\omega}{c}\sqrt{B} ,
\end{eqnarray}   
and then the $\alpha^{(m)}$ coefficients reduce to
$\alpha^{(1)} = 0 $ and $\alpha^{(2)} = (B-A)/C_1$.
Performing now some straightforward calculations we get, up to $O(\delta^2)$,
\begin{eqnarray}
\label{RefCoef1stOrder-1}
&&r_2^{\rm TE,TE}(\omega) =  r_{2_{\rm uni}}^{\rm TE,TE}(\omega) + \delta \, r_{2,1}^{\rm TE,TE}(\omega), \nonumber \\
&&r_2^{\rm TM,TE}(\omega) = - \frac{\delta \epsilon_{xx} \mu_{xx} q_{\rm tm} q_{\rm in} \, \sin 2\varphi}{(q_{\rm tm}+q_{\rm te})(q_{\rm tm} + \epsilon_{xx} q_{\rm in})(q_{\rm te} + \mu_{xx} q_{\rm in})}, \nonumber \\
&&r_2^{\rm TE,TM}(\omega) = - \frac{\delta \epsilon_{xx} \mu_{xx} q_{\rm in} \, \sin 2\varphi}{(q_{\rm tm} + \epsilon_{xx} q_{\rm in})(q_{\rm te} + \mu_{xx} q_{\rm in})}, \nonumber \\
&&r_2^{\rm TM,TM}(\omega) =  r_{2_{\rm uni}}^{\rm TM,TM}(\omega) + \delta \, r_{2,1}^{\rm TM,TM}(\omega), 
\end{eqnarray} 
where $r_{2_{\rm uni}}^{\rm TE,TE}$, $r_{2_{\rm uni}}^{\rm TM,TM}$ are given respectively by (\ref{AnisotropicFresnelCoefficients1}), (\ref{AnisotropicFresnelCoefficients2}), and we have also
defined
\begin{eqnarray}
&&q_{\rm te} \equiv  \sqrt{\epsilon_{xx}\mu_{xx} \frac{\omega^2}{c^2} - \mu_{xx}k^2},
\nonumber\\
&&q_{\rm tm} \equiv  \sqrt{\epsilon_{xx}\mu_{xx} \frac{\omega^2}{c^2} - \frac{\epsilon_{xx}}{\epsilon_{zz}}k^2},\nonumber \\
&&r_{2,1}^{\rm TE,TE}  \equiv - 
\frac{(\omega^2/2 c^2) \epsilon_{xx}  \mu_{xx} \cos^2 \!\varphi}{q_{\rm te}(q_{\rm te} + \mu_{xx} q_{\rm in})} \left(1 +  r_{2_{\rm uni}}^{\rm TE,TE} \right) , \nonumber \\
&&r_{2,1}^{\rm TM,TM} \equiv 
\frac{ (q_{\rm in}/2)\epsilon_{xx} \sin^2 \!\varphi}{q_{\rm tm} + \epsilon_{xx} q_{\rm in}}  \left(1 -  r_{2_{\rm uni}}^{\rm TM,TM} \right) . \nonumber
\end{eqnarray}

Now let us return to the general structure of the Lifshitz formula. Since ${\bf R}_2$ is not diagonal we should expect contributions coming from the non-diagonal terms in (\ref{EnergyAnisotropic}), but it can be shown that they are all $O(\delta^2)$, and therefore can be dropped. After a few rearrangements we arrive at our final expression for the Casimir pressure:
\begin{eqnarray}
\label{CasimirForce1stOrder}
&&\frac{F}{A} = 2 \hbar \int_0^{\infty}  \frac{d\xi}{2 \pi} 
\int  \frac{d^2 {\bf k}_{\|}}{(2 \pi)^2} \, K_3  \; \sum_{p={\rm TE,TM}}
\left[ I_{\rm uni} \right. \nonumber \\ && \left. +
\delta \, \frac{ r^{\rm p,p}_1 r^{\rm p,p}_{2,1} e^{-2 K_3 d} }{1- r^{\rm p,p}_1 r^{\rm p,p}_{2_{\rm uni}} e^{-2K_3 d}}
(1 + I_{\rm uni})  + O(\delta^2) \right],
\end{eqnarray}
where
\begin{eqnarray}
I_{\rm uni} = \frac{ r^{\rm p,p}_1 r^{\rm p,p}_{2_{\rm uni}} e^{-2 K_3 d} }{1- r^{\rm p,p}_1 r^{\rm p,p}_{2_{\rm uni}} e^{-2K_3 d}} .
\end{eqnarray}
An easy consistency check of this result is to take the zero-anisotropy limit, which reduces immediately to the uniaxial result, as it should. A less trivial result is to obtain the non-retarded limit of expression (\ref{CasimirForce1stOrder}), which can be shown to be consistent at first order with other results in the literature \cite{BarashAnisotropy, ParsegianWeiss, LeonhardtAnisotropic}.

Let us now assume that all the in-plane anisotropy is coded in the filling factors, just like the out-of-plane anisotropy was in the uniaxial case. In Fig. 6 we show the effects of  a slight in-plane anisotropy on the Casimir force. Each band in the plot corresponds to a different value of $f_x$, and its width is given by a $\pm 20\%$ variation of $f_y$ around $f_x$. We see that the anisotropy effect is more pronounced at small distances, because in the non-retarded limit the contribution of the electric response to the Casimir force is maximized.

\begin{center}
\begin{figure}[t]
\scalebox{0.38}{\includegraphics{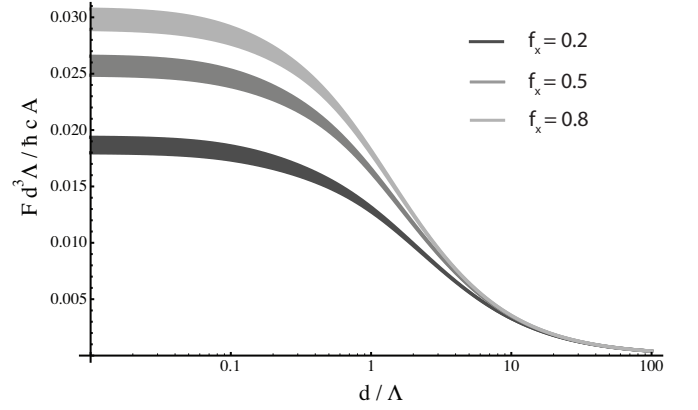}}  
\caption{The Casimir force between a gold half-space and an orthorhombic, slightly in-plane anisotropic MM for different values of the filling factors $f_x$ and $f_y$. The bands are characterized by a certain value of $f_x$, as shown in the legend, and a continuum of values of $f_y$, from $f_y = 0.8 f_x$ to $f_y = 1.2 f_x$. All the other parameters involved are exactly the same as those used in Fig.4.}
\label{Figure6}
\end{figure}
\end{center}

%%%%%%

\subsection{Dissipation effects}

\begin{center}
\begin{figure}[t]
\scalebox{0.33}{\includegraphics{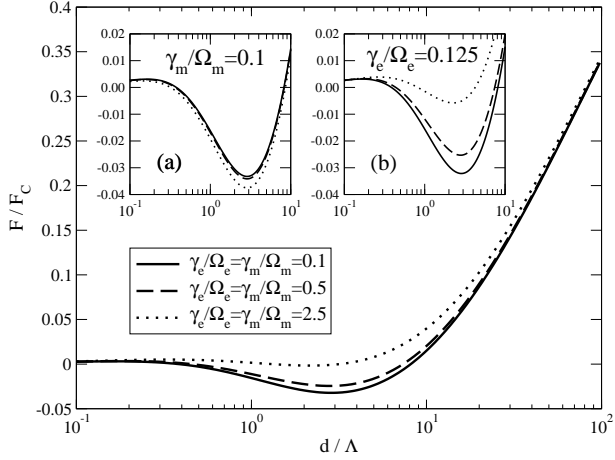}}   \\ \vspace{0.2in}
\caption{The ratio $F/F_{\rm C}$ between a gold half-space and an isotropic silver-based metamaterial 
for different values of the dissipation parameters. The main plot shows the effect
of the simultaneous modification of  electric and magnetic dissipation. Inset (a)
shows the effect of electric dissipation alone for different values of the
ratio $\gamma_e/\Omega_e$= 0.1 (solid), 0.5 (dashed), 2.5 (dotted).  
Inset (b) shows the effect of magnetic dissipation alone for different values
of the ratio $\gamma_m/\Omega_m$= 0.1 (solid), 0.5 (dashed), 2.5 (dotted).
The filling factor is $f=10^{-4}$ in all three plots, and all other parameters except the dissipation coefficients are the same as in Fig. 4.}
\label{Fig7}
\end{figure}
\end{center}

Let us now turn from considerations of anisotropy to other practical issues for MMs and Casimir interactions. It is known that dissipation plays an important role in metallic-based metamaterials, especially
those operating at high frequencies. In Fig. \ref{Fig7} we show the effect of a simultaneous modification in the electric and magnetic dissipation coefficients; it may be clearly seen that an equal change in the rates $\gamma_e / \Omega_e$ and $\gamma_m / \Omega_m$ favors attraction. In the insets (a) and (b) we show respectively the effects of changing only the electric and magnetic dissipation, that may be straightforwardly interpreted in light of the discussion presented in \cite{klich}. From (\ref{Drude Background}) we see that an increase in $\gamma_{e}$ makes $\epsilon$ smaller, pushing the metamaterial slightly closer to the Boyer limit (that is, $\epsilon_1 \rightarrow \infty$, $\mu_1 = 1$, $\epsilon_2 = 1$, $\mu_2 \rightarrow \infty$). We should thus expect an increase in the Casimir repulsion as we make $\gamma_e$ larger, and that is exactly what is observed in the inset (a).  A similar reasoning may be applied to inset (b), but since this time we are going away from the Boyer limit, repulsion diminishes as we increase $\gamma_m$. 

%\begin{widetext}
%\end{widetext}
 
%%%%%%%%%%

\subsection{Temperature effects}

The effects of temperature in the Casimir force between metamaterials and dielectrics were thoroughly discussed in \cite{henkel}, where the authors show that in this case temperature works against repulsion, or in other words, that for sufficiently high temperatures repulsion is completely overturned into attraction. In this section, we wish to extend that discussion for a metallic plate facing a MM and compare the situations where the metal is modeled by either Drude or plasma permittivities. 

In Fig. 8a we show the Casimir force for different temperatures between a Drude metal and an isotropic MM with no Drude background. In this case we see that temperature also works against repulsion, but in such a way that keeps the repulsion window quite open for temperatures as high as $T = 600 {\rm K}$, allowing for repulsion at room temperature, at least in principle. Something even more interesting happens when we change the Drude metal by a plasma metal (i.e., vanishing relaxation parameter $\gamma_1 = 0$ in eq. (\ref{Drude})), as shown in Fig. 8b. In this case, we see that not only a temperature increase does not switch back the force into attraction for large distances, but it actually increases repulsion in that regime. It is possible to explain this phenomenon in simple terms using the Lifshitz formula. Let us consider the force between two isotropic materials at a finite temperature $T$
\begin{eqnarray}
\label{LFT}
\frac{F(d,\beta)}{A}=  \hspace{-8pt}&& \frac{\hbar}{\beta} {\sum_{n=0}^{\infty}}' \sum_{p=\rm TE, TM}  \int  \frac{d^2 {\bf k}_{\|}}{(2 \pi)^2}
\, K_3 \nonumber \\
&&\times \, \frac{ r_1^{p,p}(\xi_n) \, r_2^{p,p}(\xi_n) \, e^{-2K_3 d} }{
1 - r_1^{p,p}(\xi_n)\, r_2^{p,p}(\xi_n) \, e^{-2K_3 d}} ,
\end{eqnarray}    
where the prime in the summation means that the $n=0$ term is multiplied by $1/2$, $\beta = 1/k_{\rm B} T$, $K_3 = \sqrt{k_{\|}^2 + \xi_n^2/c^2}$, $\xi_n = 2 \pi n / \hbar \beta$ are the Matsubara frequencies, and the reflection coefficients are given by (\ref{FresnelCoefficients}) and (\ref{FresnelCoefficientsMM}) with $\xi_n$ instead of $\xi$. From (\ref{LFT}) we see that for large distances (provided that $k_B T d / \hbar c \gg 1$) the $n=0$ dominates all the others, and we may approximate the Casimir force by
\begin{eqnarray}
\label{LFTzm}
\frac{F(d,\beta)}{A} =\hspace{-8pt}&& \frac{\hbar}{4\pi\beta} \sum_{p= \rm TE, TM}  \int  dk \, k^2 \nonumber \\
&& \times \, \frac{ r_1^{p,p}(0,k) r_2^{p,p}(0, k) \, e^{-2 k d} }{
1 - r_1^{p,p}(0,k)\, r_2^{p,p}(0,k) \, e^{-2 k d}} ,
\end{eqnarray}    
where $k = \vert {\bf k}_{\|} \vert$ and the reflection coefficients are evaluated at the zeroth Matsubara frequency $\xi_0 = 0$. The key difference from the setups using dieletrics or Drude metals to the one with plasma metals is that in the former cases we have $\lim_{\xi \rightarrow 0} \epsilon(\xi) \xi^2/c^2 = 0$, leading to
\begin{eqnarray}
r_1^{\rm TE,TE}(0,k) =  \frac{k - k}{k + k} = 0 ,
\end{eqnarray}
while in the latter we have
\begin{eqnarray}
r_1^{\rm TE,TE}(0,k) = \frac{k - \sqrt{k^2 + \Omega_{1}^2}}{k + \sqrt{k^2 + \Omega_{1}^2}} \leq 0 .
\end{eqnarray}
This means that for dielectrics or Drude metals facing a MM, the only contribution to (\ref{LFTzm}) comes from the TM zero mode, which is always positive (given that $r_1^{\rm TM,TM}(0,k) r_2^{\rm TM,TM}(0,k) > 0$). Since this term dominates for large distances, we conclude then that the force is attractive in this regime. However, for plasma metals facing a MM we see that both TE and TM zero modes contribute, and while $r_1^{\rm TM,TM}(0,k) r_2^{\rm TM,TM}(0,k)$ is positive the product $r_1^{\rm TE,TE}(0,k) r_2^{\rm TE,TE}(0,k)$ is not, due to the different signs of $r_1^{\rm TE,TE}(0,k) \leq 0$ and of $r_2^{\rm TE,TE}(0,k)$  
\begin{eqnarray}
r_2^{\rm TE,TE}(0,k) = \frac{\mu_2(0) - 1}{\mu_2(0) + 1} > 0 .
\end{eqnarray}
We see then that the sign of the force depends on a delicate balance between the TE and TM contributions, and it so happens that for our chosen parameters the TE term overwhelms the TM term and repulsion is sustained for all distances above the crossover from attraction. The fact that repulsion is enhanced is also easily explained, since a simple analysis shows that for large distances (\ref{LFTzm}) may be put in the form $C / \beta d^3$, where $C$ is a constant depending on the materials used. It is clear then that if $C$ is negative, a temperature increase can only enhance repulsion.  Our findings for temperature effects using either Drude or plasma models for the metallic plate are consistent with the conclusions of \cite{MostPRL} .
\begin{center}
\begin{figure}[t]
\subfigure{\scalebox{0.4}{\includegraphics{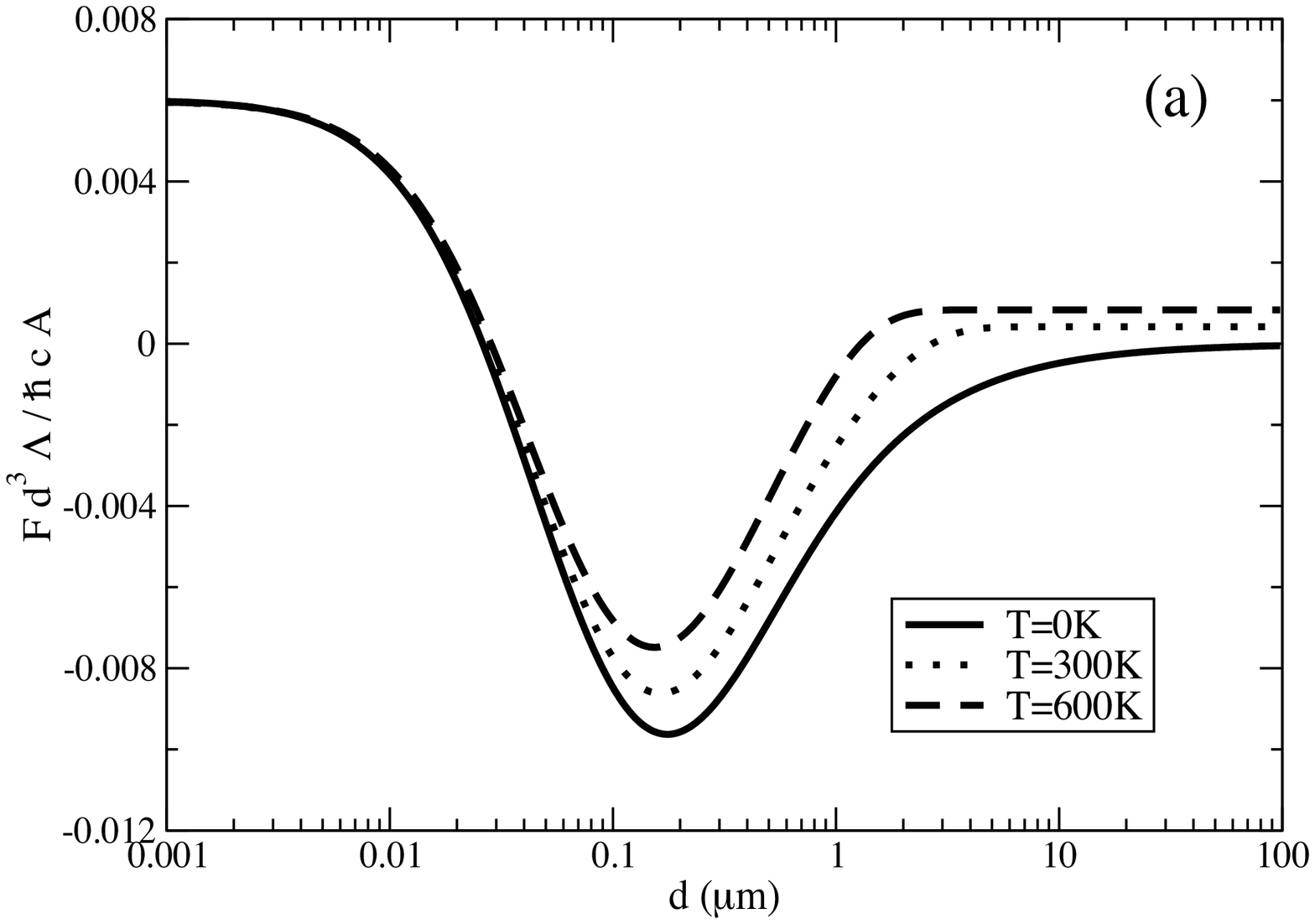}}}   \\  
\vspace{-25pt}\subfigure{\scalebox{0.4}{\includegraphics{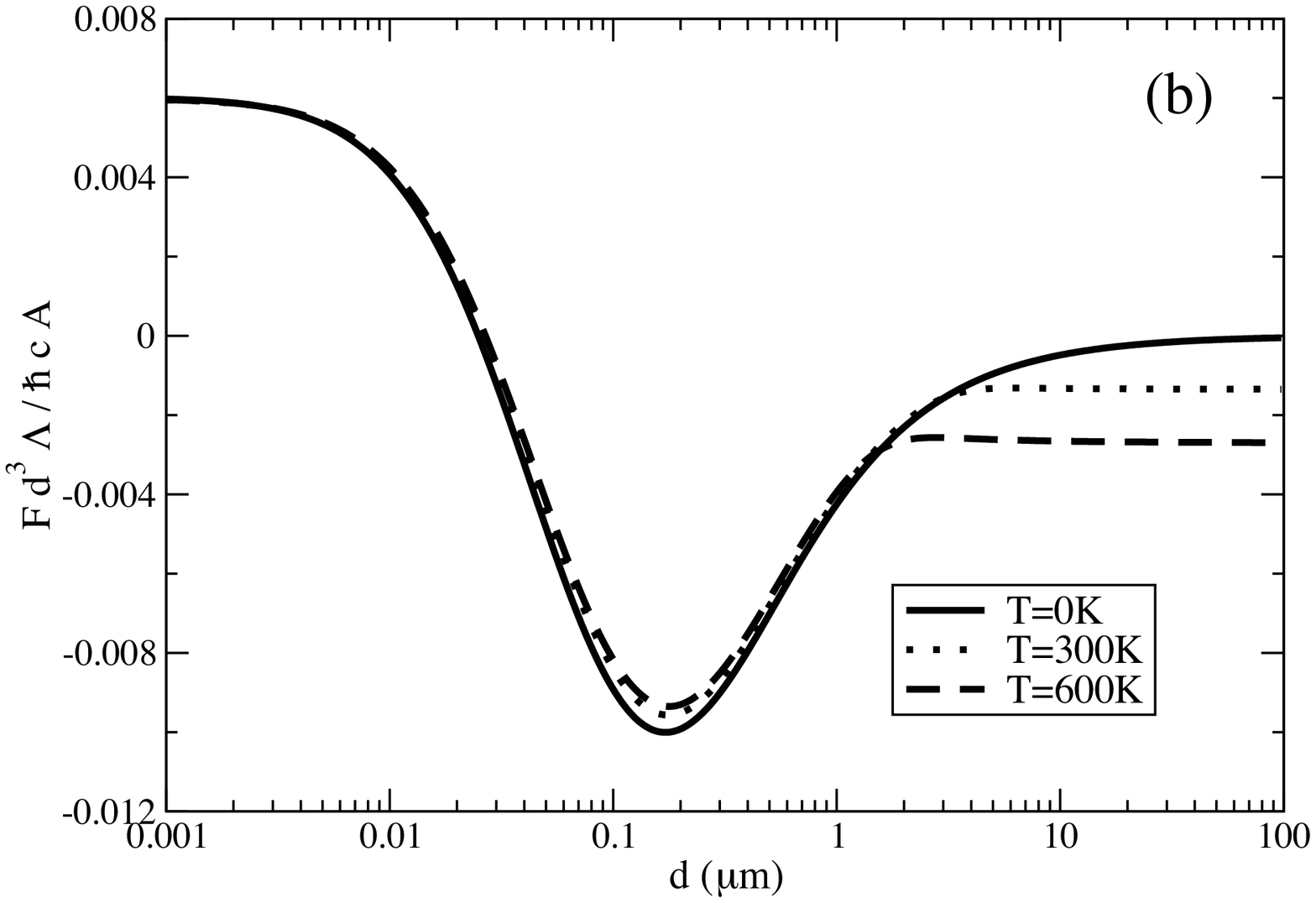}}}  
\caption{Temperature dependence of the Casimir force between a metallic plate and a metamaterial.  
We plot the Casimir force between a metamaterial and a Drude metal (a) or a plasma metal (b) for different temperatures. We stress that negative values of the force characterize repulsion, and that all parameters are the same as the ones used in the $f=0$ curve of Fig. 4.}
\label{Fig8}
\end{figure}
\end{center}

\subsection{Metamaterials based on isolated metallic structures}

There are several examples of metamaterials where the metallic part is distributed in a non-connected way. In \cite{grigorenko}, for instance, the authors put forward a MM consisting of pairs of metallic nanopillars, regularly distributed on top of a dielectric substrate. The pairing of pillars is necessary to
create an antisymmetric resonance (when the currents in each pillar are running in opposite directions) at a certain frequency, where the electric dipole contributions of both pillars are nearly canceled out and the effective current loops produced by the pairs give rise to magnetic dipole contributions, resulting in a non-trivial magnetic activity.

Unfortunately, a detailed treatment of the metamaterial previously described is beyond the scope of the present paper. It is still possible, however, to capture some effects of metallic non-connectedness and geometrically built-in resonances through the use of an appropriate toy model. 
In order to address the first issue, we consider a simple MM model consisting of identical, small metallic spheres of radius $a$ regularly distributed in a host dielectric (non-magnetic) medium. Assuming that the metal and the dielectric are characterized respectively by the permittivities 
\begin{eqnarray}
\label{Emet}
&&\epsilon_{2, {\rm met}}(\omega) =  1 - \frac{\Omega_{2,{\rm met}}^2}{\omega^2 + i \gamma_{2,{\rm met}} \omega} \nonumber \\
\label{Ediel}
&&\epsilon_{2, {\rm d}}(\omega) = 1 - \sum_{i=1}^{N} \frac{\Omega_{2,i}^2}{\omega^2 - \omega_{2,i}^2 + i \gamma_{2,i} \omega} \, ,
\end{eqnarray}
and that the metallic spheres can be considered in a first approximation as electric dipoles, one can connect the medium effective permittivity $\epsilon_{2,{\rm nc}}(\omega)$ to the electric polarizability $\alpha(\omega)$ of a given sphere through the Clausius-Mossotti formula \cite{Ruppin}
\begin{eqnarray}
\label{CM1}
\frac{f}{a^3} \alpha = \frac{\epsilon_{2,{\rm nc}}  - \epsilon_{2, {\rm d}}}{\epsilon_{2,{\rm nc}}  + 2\epsilon_{2, {\rm d}}} \, ,
\end{eqnarray}
where $f$ is the metallic filling factor and we have suppressed the $\omega$-dependence for simplicity. It is also possible to show that when $(\omega/c)a \ll 1$ (the spheres are much smaller than the radiation wavelength), the electric polarizability may be given in terms of the dielectric function of the metal by the similar relation
\begin{eqnarray}
\frac{\alpha}{a^3} = \frac{\epsilon_{2,{\rm met}} - \epsilon_{2, {\rm d}}}{\epsilon_{2,{\rm met}} + 2\epsilon_{2, {\rm d}}} \, ,
\end{eqnarray}
and by eliminating $\alpha$ in (\ref{CM1}) we get
\begin{eqnarray}
\label{permittivityMG}
\epsilon_{2,\rm nc}(\omega) = \epsilon_{2,d} \frac{(1+2f)\epsilon_{2,{\rm met}} + 2(1-f)\epsilon_{2, {\rm d}}}{(1-f)\epsilon_{2,{\rm met}} + (2+f)\epsilon_{2, {\rm d}}} \, .
\end{eqnarray} 
This result is known as the Maxwell Garnett approximation for the permittivity \cite{Yannopapas,Ruppin}, after the physicist who derived it in the early 1900s \cite{MaxwellGarnett}. A brief analysis shows the main effect of having isolated metallic pieces: the previous formula tends to a finite value in the zero frequency limit, unlike (\ref{Drude Background}), which describes a connected metallic MM. The effective permeability can be dealt in a similar way, and it is possible to show that in this approximation we have simply $\mu_{2,\rm nc}(\omega) = 1$. 

As noted earlier, formula (\ref{permittivityMG}) accounts only for effects of metallic non-connectedness. In order to include the built-in electric and magnetic resonances \cite{foot1}, we simply assume their existence in an {\it ad hoc} manner and add their contribution to $\epsilon_{2,\rm nc}(\omega)$ and $\mu_{2,\rm nc}(\omega)$, respectively. Assuming those resonances can be modeled by Drude-Lorentz formulas, we have, finally,
\begin{eqnarray}
\label{EpsilonMuNC}
\epsilon_{2}(\omega) \hspace{-6pt}&&= \epsilon_{2,{\rm b}}(\omega) + \epsilon_{2, {\rm res}}(\omega) \nonumber \\
&&= \frac{(1+2f)\epsilon_{2, {\rm met}} + 2(1-f)\epsilon_{2, {\rm d}}}{(1-f)\epsilon_{2, {\rm met}} + (2+f)\epsilon_{2, {\rm d}}} \nonumber \\
&&+\,  \frac{\Omega_{\rm e}^2}{\omega^2 - \omega_{\rm e}^2 + i \gamma_{\rm e} \omega} \nonumber \\
\mu_2(\omega) \hspace{-6pt}&&= 1 - \frac{\Omega_{\rm m}^2}{ \omega^2 - 
\omega_{\rm m}^2 + i \gamma_{\rm m} \omega} \, .
\end{eqnarray}
The results for the Casimir force are shown in Fig. 9. The parameters for the resonant parts $\epsilon_{2, {\rm res}}(\omega)$ and $\mu_{2, {\rm res}}(\omega)$ are roughly based on the experimental results given in \cite{grigorenko} for a MM consisting of metallic nanopillars covered with a thin layer of glycerine. As indicated earlier, our intention here is not to provide a precise description of such experiments, but only to estimate how this type of metamaterial affects the Casimir force. The embedding dielectric, glass BK7, is quite well described by (\ref{Ediel}) with the parameters $N=3$, $\Omega_{2,1}/\Omega = 1.84$, $\omega_{2,1}/\Omega = 1.81$, $\Omega_{2,2}/\Omega = 0.47$, $\omega_{2,2}/\Omega = 0.28$, $\Omega_{2,3}/\Omega = \omega_{2,3}/\Omega = 0.014$, $\gamma_{2,1}/\Omega = \gamma_{2,2}/\Omega = \gamma_{2,3}/\Omega = 0$. It is clearly seen that no repulsion is achieved, and the reason is that the magnetic resonance created by the MM geometry is too weak to overwhelm the electric background. In other words, the MM is mainly dielectric, leading to an attractive force. 
\begin{center}
\begin{figure}[t]
\scalebox{0.32}{\includegraphics{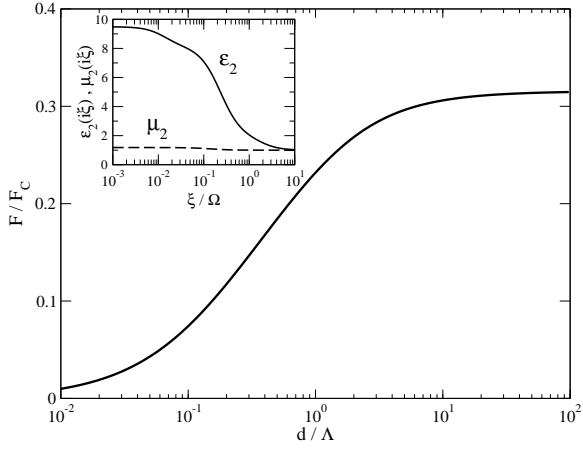}}
\caption{The ratio $F/F_{\rm C}$ for a gold half-space facing a isotropic, non-connected and gold-based metamaterial. The parameters for the metal are $\Omega_{2,{\rm met}}/\Omega=0.96$, $\gamma_{2,{\rm met}}/\Omega=0.004$, and for the metamaterial we have $\Omega_{\rm e}/\Omega=0.34$, $\Omega_{\rm m}/\Omega=0.064$, $\omega_{\rm e}/\Omega= 0.2$, $\omega_{\rm m}/\Omega=0.15$, $\gamma_{\rm e}/\Omega=0.04$, $\gamma_{\rm m}/\Omega=0.02$, $f=0.1$. The inset shows the permittivity and permeability inside the MM, as given by (\ref{EpsilonMuNC}), but as functions of imaginary frequencies $\xi$.}
\label{fig5}
\end{figure} 
\end{center}

%%%%%%%%%%%%

\section{Dielectric-based metamaterials and the Casimir effect}

Metamaterials based exclusively on dielectrics \cite{Wheeler,Huang,Schuller} are an interesting alternative to metallic MMs. For one thing, they provide new possibilities for the construction of negative index materials \cite{foot2}, since they allow for both the permittivity and the permeability to assume negative values in bandwidths that may be out of reach with metallic-based MMs. In addition, dielectric-based MMs might be interesting for Casimir force studies for the same reason that non-connected metallic MMs might be: they do not present a Drude background at low frequencies, and this is
advantageous for the observation of magnetic effects in the Casimir force.

The dielectrics most commonly used in the construction of MMs are ``polaritonic" crystals \cite{MillsBurstein} characterized by the dielectric function         
\begin{eqnarray}
\label{polaritonic}
\epsilon_{\rm pol}(\omega) = \epsilon_{\infty} \left(1 + \frac{\Omega_{\rm pol}^2 - \omega_{\rm pol}^2 }{ -\omega^2 + \omega_{\rm pol}^2 + i \gamma_{\rm pol}\omega} \right) ,
\end{eqnarray}
where $\omega_{\rm pol}$ is a characteristic resonance of the system, $\epsilon_{\infty}$ is the permittivity at very high frequencies, and $\Omega_{\rm pol}  = \omega_{\rm pol} \sqrt{\epsilon(0)/\epsilon_{\infty}}$. In order to fix ideas, let us consider a MM made of a regular array of polaritonic nanospheres of radius $a$ embedded in an isotropic dielectric and non-magnetic host characterized by a dielectric function $\epsilon_{h}$. For sufficiently long wavelengths and sparse arrays, meaning $x \equiv \omega R / c \ll 1$, it is possible to use the so-called extended Maxwell-Garnett theory \cite{Yannopapas} to evaluate the dielectric and magnetic properties of the metamaterial, giving \cite{Yannopapas, Wheeler}
%
%This method is founded on the Mie theory of scattering \cite{BornWolf}, from where  
%and subsequent truncation at the dipole contributions. The use of Lorentz-Lorenz formula then leads to the following effective permittivity and permeability functions
%
\begin{eqnarray}
\label{MGextended}
\epsilon_{\rm emg}(\omega) = \epsilon_h \frac{x^3 - 3i f a_1}{x^3 + \frac{3}{2}i f a_1} \;\; , \;\;  \mu_{\rm emg}(\omega) =  \frac{x^3 - 3i f b_1}{x^3 + \frac{3}{2}i f b_1} 
\end{eqnarray}
where $f$ is the array filling factor and $a_1$, $b_1$ are respectively the electric and magnetic dipole coefficients of the scattering matrix of a single sphere, given by \cite{BornWolf, BohrenHuffman}
\begin{eqnarray}
\label{definitionsMGextended}
&&a_1 = \frac{j_1(x_{\rm pol}) [x j_1(x)]' \epsilon_{\rm pol} - j_1(x) [x_{\rm pol} j_1(x_{\rm pol})]' \epsilon_h}{h^{(+)}_1(x) [x_{\rm pol} j_1(x_{\rm pol})]' \epsilon_h - j_1(x_{\rm pol}) [x h^{(+)}_1(x)]' \epsilon_{\rm pol}} \nonumber \\
&&b_1 = \frac{j_1(x_{\rm pol}) [x j_1(x)]'  - j_1(x) [x_{\rm pol} j_1(x_{\rm pol})]'}{h^{(+)}_1(x) [x_{\rm pol} j_1(x_{\rm pol})]' - j_1(x_{\rm pol}) [x h^{(+)}_1(x)]'}
\end{eqnarray}
where $j_1 (h^{+}_1)$ is the spherical Bessel function (Hankel function of the first kind) of order one, $x_{\rm pol} = \sqrt{\epsilon_{\rm pol}}x$ and the prime has the usual meaning of a derivative with respect to the function argument. The important thing to notice here is the fact that $\mu_{\rm emg}$ may present several resonances even when the nanospheres are purely dielectric, from which we conclude that in this framework we do not have to assume an {\it ad hoc} resonant behavior; it is already built into the theory.  

%\begin{widetext}
\begin{center}
\begin{figure}
{\scalebox{0.32}{\includegraphics{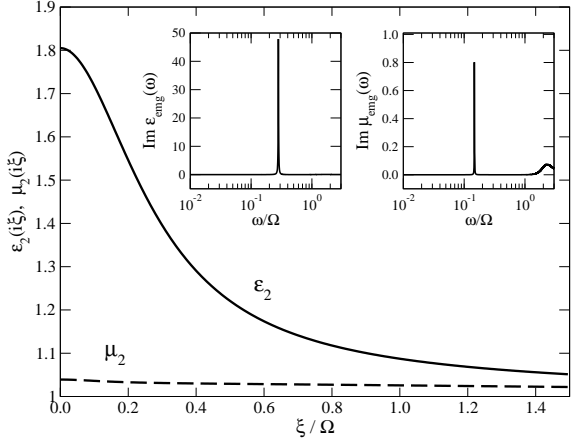}}
\caption{The permittivity $\epsilon_{\rm true}(i \xi)$ and permeability $\mu_{\rm true}(i \xi)$. The parameters are $\epsilon_{\infty} =2$, $\Omega_{pol}/\Omega = 0.4$, $\omega_{pol}/\Omega = 0.15$, $\gamma_{pol}/\Omega = 0.001$. }
\label{fig8}}
\end{figure} 
\end{center}
%\end{widetext}

The usual procedure at this point would be to rotate expressions (\ref{MGextended}) to the imaginary frequency axis and substitute them into the appropriate reflections coefficients, but in this case things are not so straightforward. 
Implicit in the Lifshitz formula for imaginary frequencies it is the assumption of analyticity of $\epsilon(\omega)$ and $\mu(\omega)$ in the upper half-plane, a condition that $\epsilon_{\rm emg}(\omega)$ and $\mu_{\rm emg}(\omega)$ do not satisfy. In order to overcome this obstacle we have to remind ourselves that expressions (\ref{MGextended}) were derived as approximations to the true permittivity  $\epsilon(\omega)$ and permeability $\mu(\omega)$ only for a given range of {\it real} frequencies, namely, for $\omega$ such as $\omega R / c \ll 1$. This means that while $\epsilon(\omega)$ and $\mu(\omega)$ must be analytic in the upper half-plane due to causality requirements, $\epsilon_{\rm emg}(\omega)$ and $\mu_{\rm emg}(\omega)$ are not necessarily bound to causal behavior. In other words, it means that the analytical continuations of  $\epsilon_{\rm emg}(\omega)$ and $\mu_{\rm emg}(\omega)$ into the complex plane are not necessarily close to the continuations of $\epsilon(\omega)$ and $\mu(\omega)$, and in this case they happen to be quite different. 

A possible way to proceed is to rely on the analytic properties of $\epsilon(\omega)$ and write the Kramers-Kronig relation \cite{LandauContMedia} 
\begin{eqnarray}
\label{KK1}
\epsilon(\omega) = 1 + \frac{1}{i\pi} {\rm P} \int_0^{\infty} dy \frac{\epsilon(y) - 1}{y - \omega} ,
\end{eqnarray}
where ${\rm P}$ stands for the Cauchy principal value, and consider also the analogous relation for $\mu(\omega)$. Taking the real part and evaluating it at an imaginary frequency $i \xi$, we obtain
\begin{eqnarray}
\label{KK2}
\epsilon(i \xi)  = 1 + \frac{2}{\pi} \int_0^{\infty} dy y \frac{ {\rm Im} \,\epsilon(y)}{\xi^2 + y^2} ,
\end{eqnarray}
and, using the fact that $\epsilon(\omega) \approx \epsilon_{\rm emg}(\omega)$ \cite{foot3}, we have 
\begin{eqnarray}
\label{EpsilonMuImEmg}
&& \epsilon(i \xi) \approx 1 + \frac{2}{\pi} \int_0^{\infty} dy y \,\frac{{\rm Im} \, \epsilon_{\rm emg}(y)}{\xi^2 + y^2} , \nonumber \\
&& \mu(i \xi) \approx 1 + \frac{2}{\pi} \int_0^{\infty} dy y\, \frac{{\rm Im} \, \mu_{\rm emg}(y)}{\xi^2 + y^2} .
\end{eqnarray}
In Fig. 10 we plot $\epsilon(i \xi)$ and $\mu(i \xi)$ using approximate values for TlCl polaritonic spheres \cite{Huang} embedded in vacuum. We see that $\epsilon(i \xi)$ is overwhelmingly dominant over $\mu(i \xi)$, which in fact is hardly different from unity. As the insets show, this is basically due to a single strong resonance, around $\omega = 0.3 \Omega$, that appears in $\epsilon_{\rm emg}(\omega)$ but not in $\mu_{\rm emg}(\omega)$. From these results we conclude that, despite the fact that some magnetic activity is created by an array of polaritonic spheres, the Casimir force in this case is dictated by the electric part alone and therefore no repulsion seems possible.

%%%%%%%%%%%%%%%%%%%%%%%%%%%%%%%%%%%%%%%%%%%%%%%%%%%
%%%%%%%%%%%%%%%%%%%%%%%%%%%%%%%%%%%%%%%%%%%%%%%%%%%

\section{Discussion}

A striking confirmation of the magnetic influence on the Casimir force would be a measurement of repulsion between a metallic plate and a magnetodielectric one. This seems unlikely in light of the examples presented here, but a measured reduction in the attractive force might nevertheless be traced back to the magnetic properties of a metamaterial.       
\begin{center}
\begin{figure}
{\scalebox{0.32}{\includegraphics{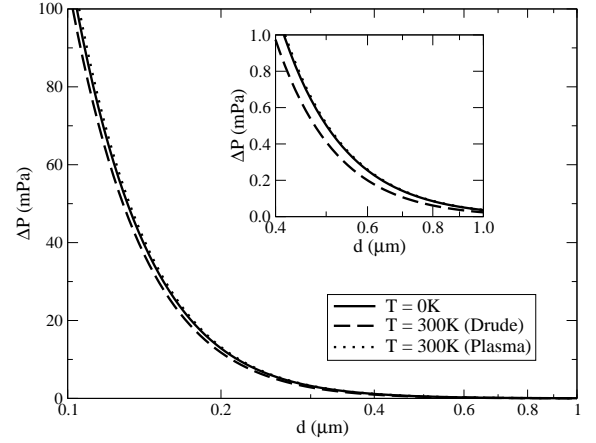}}
\caption{The plot of $\Delta P = P^{(2)} - P^{(1)}$ for different temperatures and models. Following our conventions throughout the paper, a positive force means attraction. The parameters of $P^{(1)}$ are the same used in Fig. 8, in dimensional units they are $\Omega_D = 1.32 \times 10^{16} {\rm rad/s}$, $\gamma_D = 5.48 \times 10^{13} {\rm rad/s}$ ($\gamma_D = 0$ for the plasma curve), $\Omega_e = 4.7 \times 10^{15} {\rm rad/s}$, $\Omega_m = 8.7 \times 10^{14} {\rm rad/s}$, $\omega_e = 2.7 \times 10^{15} {\rm rad/s}$, $\omega_m = 2 \times 10^{15} {\rm rad/s}$, $\gamma_e/ = 5.5 \times 10^{14} {\rm rad/s}$, $\gamma_m = 2.7 \times 10^{14} {\rm rad/s}$, $\Omega_{2,1} = 2.52 \times 10^{16} {\rm rad/s}$, $\omega_{2,1} = 2.48 \times 10^{16} {\rm rad/s}$, $\Omega_{2,2} = 6.4 \times 10^{15} {\rm rad/s}$, $\omega_{2,2} = 3.8 \times 10^{15} {\rm rad/s}$, $\Omega_{2,3} = \omega_{2,3} = 1.9 \times 10^{14} {\rm rad/s}$, $\gamma_{2,1} = \gamma_{2,2} = \gamma_{2,3} = 0$, $f=0.1$, and the parameters of $P^{(2)}$ are exactly the same except for $\Omega_e = 0$. The inset shows the same plot on a different scale, since in the larger one it is not possible to see $\Delta P$ for large distances.}
\label{fig10}}
\end{figure} 
\end{center}

Let $P^{(1)}$ be the Casimir pressure between a gold half-space and a given metamaterial.  If the
magnetic properties of the MM are  {\lq\lq turned off\rq\rq}, keeping all other parameters the same, the
pressure will change to some new value $P^{(2)}$. In order to check whether the difference $\Delta P = P^{(1)} - P^{(2)}$ should be observable, we plot its computed value in Fig. 11 for zero and room temperatures, using both Drude and plasma models for the metal. The sensitivity of current experiments lies around $1$ mPa, from which we conclude that detection of magnetic effects in our setup is currently possible up to $d \sim 0.4 {\rm \mu m}$. While this suggests a considerably large window for measurement, given that many experiments probe the $150 - 350$ $\mu$m range quite accurately, several things must be dealt with. First and foremost, we see that the difference between the Drude and plasma predictions are considerably large (as compared to the magnetic effect) above $0.6 \mu m$. This means that in order to ascribe changes in the Casimir force ambiguously to magnetic effects one has to know how to model metallic materials properly. In addition, at close distances like $d \lesssim 0.4 \mu m$, the effective medium approximation probably no longer holds, since the very structures that produce magnetic activity (the metallic spheres in this example) are built on the scale of hundreds of nanometers or larger. These finite-size effects should bring significant corrections to the Casimir force, and must be considered in a more sophisticated analysis. Finally, there are the imperfections of the materials themselves, like roughness, that at those distances play a non-negligible role. We conclude then that despite the fact that current experiments have in principle the sensitivity necessary to detect magnetic effects, an actual measurement of such effects remains a challenging task.    

\begin{center}
\begin{figure}
{\scalebox{0.32}{\includegraphics{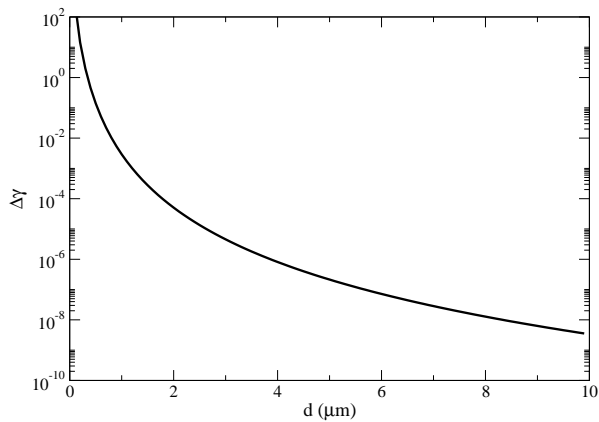}}
\caption{The difference in frequency shifts caused by the presence or absence of magnetic activity in the metamaterial. Everything is assumed to be at zero temperature. The MM is the same used in Fig. 11, and the parameters for the Rb atom are $m = 1.45 \times 10^{-25}$ Kg, $\alpha_0 = 4.74 \times 10^{-23} {\rm cm}^3$ and  $\omega_0 = 2.54 \times 10^{15}$ rad/s, with an unperturbed trap frequency of $\omega_z = 2\pi \times 229$ Hz.}
\label{fig11}}
\end{figure} 
\end{center}

Casimir-Polder experiments \cite{CasimirPolder, Cornell} also provide possibilities for the detection of magnetic effects. These experiments are able to probe larger distances than the typical bulk-bulk measurements, which is desirable from the point of view of an effective medium approximation. The zero temperature Casimir-Polder potential between a ground state atom and a material half-space is \cite{Buhmann},
\begin{eqnarray}
\hspace{-10pt}&&U_{\rm CP}(z) = \frac{\hbar}{8 \pi^2 c} \int_0^{\infty} d\xi \xi^2 \alpha(i\xi)  \int_0^{\infty} dk k \frac{e^{-2z K_3}}{K_3} \nonumber \\
&& \times \left[ r^{\rm TE,TE}(i\xi,k) - \left(1 + \frac{2k^2c^2}{\xi^2} \right) r^{\rm TM,TM}(i\xi,k) \right] , \nonumber \\
\end{eqnarray} 
where $z$ is the distance between the atom and the half-space, $K_3$ is defined just below (\ref{eq1}), and 
$r^{\rm TE,TE}$ and $r^{\rm TM,TM}$ are given by (\ref{FresnelCoefficients}). $\alpha(i \xi)$ is the dynamic atomic polarizability, which we assume is described reasonably well by the single-resonance expression
\begin{eqnarray}
\alpha(i \xi) = \frac{\alpha_0}{1 + \xi^2/\omega_0^2} ,
\end{eqnarray}
where $\alpha_0$ is the static polarizability and $\omega_0$ is the dominant atomic transition. In one type of experiment \cite{Cornell} the directly measured quantity is the frequency shift in the center of mass oscillation of a Bose-Einstein condensate:
\begin{eqnarray}
\gamma(z) = \frac{1}{2m \omega_z^2} \left. \partial_{z'}^2 U_{\rm CP}(z') \right|_{z'=z} ,
\end{eqnarray} 
where $m$ is the atomic mass and $\omega_z$ is the unperturbed ({\it i.e.}, without Casimir-Polder forces) oscillation frequency. The reported sensitivity for $\gamma$ lies between $10^{-5}$ and $10^{-4}$, setting the lower bound for the detection of magnetic effects in the Casimir force. Let us then consider a Rb atom in front of the same MM used in the previous example, and compare the frequency shifts when its magnetic part is {\lq\lq turned on\rq\rq} and {\lq\lq off\rq\rq}. In Fig. 12 we plot the difference $\Delta \gamma(z) = \gamma_{nm}(z) - \gamma_{m}(z)$, where $\gamma_{m}(z)$ and $\gamma_{nm}(z)$ are respectively the frequency shifts when the magnetic activity is present and absent. We see that in the best case scenario (sensitivity equal to $10^{-5}$) the magnetic influence would be detectable up to around 2.5 $\mu$m; for larger distances the force is just too weak. 

As a final remark we note that, while Casimir repulsion will likely be very difficult to observe with existing 
metamaterials, the detection of magnetic effects through a slight reduction in the Casimir attraction is definitely possible. There are still some issues to be dealt with, like 
the assurance that magnetic activity is the main cause of force reduction, rather than some trivial effect like a reduced filling factor. With the consistent development of both Casimir measurements and MMs manufactured in the recent years, it is very reasonable to expect that a Casimir measurement of magnetic effects will be feasible in the near future.

%%%%%%%%%%

\section{Acknowledgments}

We are greatly indebted to H.-T. Chen, R.S. Decca, N. Engheta, S.K. Lamoreaux, P.A.M. Neto, J.F. O'Hara, W.J. Padilla, J.B. Pendry, V.M. Shalaev, D.R. Smith and A.J. Taylor, for very useful discussions. We also acknowledge the support of the U.S. Department of Energy through the LANL/LDRD program for this work.

%%%%%%%%%


\begin{thebibliography}{99} 

\bibitem{reviewsMM} Reviews of work on metamaterials have been given, for instance, by S.A. Ramakrishna, Rep. Prog. Phys. {\bf 2005}, {\em 68}, 449 (2005) and by V.M. Shalaev and A. Boardman, Focus Issue on Metamaterials J. Opt. Soc. Am. B {\bf 23}, 386 (2006).

\bibitem{negative_refraction} The possibility of negative refraction and some of its consequences was predicted by V.G. Veselago, Sov. Phys. Solid State {\bf 8}, 2854 (1967). It was experimentally confirmed in D.R. Smith, W.J. Padilla, D.C. Vier, S.C. Nemat-Nasser, and S. Schultz, Phys. Rev. Lett. {\bf 84}, 4184 (2000); R.A. Shelby, D.R. Smith, and S. Schultz, Science {\bf 292}, 77 (2001).

\bibitem{perfect_lens} J.B. Pendry, Phys. Rev. Lett. {\bf 85}, 3966 (2000); 
I.A. Larkin and M.I. Stockman, Nano. Lett. {\bf 5}, 339 (2005).

\bibitem{cloaking} The possibility of cloaking of objects with metamaterials was considered by
J.B. Smith, D. Schurig, and D.R.~Smith, Science {\bf 312}, 1780 (2006), and by U. Leonhardt, Science {\bf 312}, 1777 (2006).
It was experimentally demonstrated for microwave frequencies by D. Schurig, J.J. Mock, B.J. Justice, S.A. Cummer, J.B. Pendry,
A.F. Starr and D.R. Smith, Science {\bf 314}, 977 (2006).

\bibitem{Casimir} H.B.G. Casimir, Proc. K. Ned. Akad. Wet. {\bf 51}, 793 (1948).

\bibitem{reviewsCasimir} 
For recent reviews, see 
M. Bordag, U. Mohideen, and V.M. Mostepanenko, Phys. Rep. {\bf 353}, 1 (2001);
K.A. Milton, J. Phys. A {\bf 24}, R209 (2004);
S.K. Lamoreaux, Rep. Prog. Phys. {\bf 68}, 201 (2005).
 
\bibitem{casimirexperiments} 
S.K. Lamoreaux, Phys. Rev. Lett. {\bf 78}, 5 (1997); U. Mohideen and A. Roy, 
Phys. Rev. Lett. {\bf 81}, 4549 (1998); H.B. Chan, V.A. Aksyuk, R.N. Kleiman, D.J. Bishop and F. Capasso, 
Science {\bf 291}, 1941 (2001); G. Bressi, G. Carugno, R. Onofrio and G. Ruoso, Phys. Rev. Lett. {\bf 88}, 041804 (2002);
R.S. Decca, D. L\'opez, E. Fischbach and D.E. Krause, Phys. Rev. Lett. {\bf 91}, 050402 (2003).

\bibitem{dielectricrepulsion} I.E. Dzyaloshinskii, E.M. Lifshitz and L.P. Pitaevskii, Usp. Fiz. Nauk {\bf 73}, 381 (1961), 
translated in Sov. Phys. Usp. {\bf 73} 153 (1961).

\bibitem{boyer} T.H. Boyer, Phys. Rev. A {\bf 9}, 2078 (1974).

\bibitem{klich} O. Kenneth, I. Klich, A. Mann and M. Revzen, Phys. Rev. Lett. {\bf 89}, 033001 (2002).

\bibitem{capassocomment} D. Iannuzzi and F. Capasso, Phys. Rev. Lett. {\bf 91}, 029101 (2003).

\bibitem{ferrites} Good examples of materials with non-trivial magnetic response ($\mu(\omega) \neq1$) at low frequencies are electric insulators with a strong magnetic response, such as ferrites and antiferromagnets. Their magnetic response, however, tails off at the infrared or
lower frequencies.  

\bibitem{grigorenko} A.N. Grigorenko, A.K. Geim, H.F. Gleeson, Y. Zhang, A.A. Firsov, I.Y. Krushchev and J. Petrovic, Nature {\bf 438}, 335 (2005).

\bibitem{shalaev} V.M. Shalaev, Nature Photonics {\bf 1}, 41 (2007).

\bibitem{dolling} G. Dolling, C. Enkrich,  M. Wegener, C.M. Sokoulis and S. Linden, Opt. Lett. {\bf 31}, 1800 (2006); G. Dolling, M. Wegener, C.M. Sokoulis and S. Linden, ibid. {\bf 32}, 53 (2007).

\bibitem{henkel} C. Henkel and K. Joulain, Europhys. Lett. {\bf 72}, 929 (2005).

\bibitem{leonhardt} U. Leonhardt and T.G. Philbin,  New J. Phys. {\bf 9}, 254 (2007). 

\bibitem{irina} I.G. Pirozhenko and A. Lambrecht, J. Phys. A: Math. Theor. {\bf 41}, 164015 (2008).

\bibitem{zhu} Y. Yang, R. Zeng, J. Xu, and S. Liu, Phys. Rev. A {\bf 77}, 015803 (2008);
Y. Yang, R. Zeng, S. Liu, H. Chen, and S. Zhu, arXiv:0803.3382.

\bibitem{lifshitz} E.M. Lifshitz, Zh. Eksp. Teor. Fiz. {\bf 29}, 94 (1955), translated in Sov. Phys. JETP {\bf 2}, 73 (1956).

\bibitem{usPRL} F.S.S. Rosa, D.A.R. Dalvit and P.W. Milonni, Phys. Rev. Lett. {\bf 100}, 183602 (2008).

\bibitem{analyticalsolutions} T. Emig, R.L. Jaffe, M. Kardar and A. Scardicchio, Phys. Rev. Lett. {\bf 96} 080403 (2006);
A. Bulgac, P. Magierski and A. Wirzba, Phys. Rev. D {\bf 73}, 025007 (2006); M. Bordag, Phys. Rev. D {\bf 73}, 125018 (2006);
D.A.R. Dalvit, F.C. Lombardo, F.D. Mazzitelli and R. Onofrio, Phys. Rev. A {\bf 74}, 020101(R) (2006). 

\bibitem{numericalsolutions} H. Gies, K. Langfeld and L. Moyaerts, J. High Energy Phys. {\bf 06}, 018 (2003); 
A. Rodriguez, M. Ibanescu, D. Iannuzzi, F. Capasso, J.D. Joannopoulos, and S.G. Johnson, Phys. Rev. Lett. {\bf 99}, 080401 (2007).

\bibitem{BalDup} R. Balian and B. Duplantier, Ann. Phys. (N.Y.) {\bf 104}, 300 (1977); {\bf 112}, 165 (1978).

\bibitem{french} A. Lambrecht, P.A. Maia Neto, and S. Reynaud, New J. Phys. {\bf 8}, 243 (2006); P.A. Maia Neto, A. Lambrecht and S. Reynaud, arXiv:0803.2444.

\bibitem{MITgroup} T. Emig, N. Graham, R.L. Jaffe, and M. Kardar, Phys. Rev. Lett. {\bf 99}, 170403 (2007); T. Emig, J. Stat. Mech: Th. Exp., Vol. 2008, P04007 (2008); T. Emig and R.L. Jaffe, J. Phys. A: Math. Theor. {\bf 41}, 164001 (2008).

\bibitem{BornWolf} M. Born and E. Wolf, {\it Principles of Optics}, 7th ed. (Cambridge University Press, Cambridge, 2005).

\bibitem{Kong} J.A. Kong, {\it Electromagnetic Wave Theory}, 2nd ed. (Wiley, 1990), Chap. 2.

\bibitem{marques} R. Marqu\'es, F. Medina, and R. Rafii-El-Idrissi, Phys. Rev. B {\bf 65}, 144440 (2002).

\bibitem{Padilla} W.J. Padilla, Optics Express {\bf 15}, 1639 (2007).

\bibitem{Chew} W.C. Chew, {\it Waves and Fields in Inhomogeneous Media}, (IEEE Press, 1995).

\bibitem{Visnovsky} S. Vi\v{s}\v{n}ovsky, {\it Optics in Magnetic Multilayers and Nanostructures}, (Taylor and Francis, Boca Raton, 2006), Chap. 2.

\bibitem{Kittel} C. Kittel, {\it Introduction to Solid State Physics}, 2nd ed. (John Wiley and Sons, New York, 1962), Chap. 1.

\bibitem{PendrySRR} J.B. Pendry, A.J. Holden, W.J. Stewart, and I. Youngs, Phys. Rev. Lett {\bf 76}, 4773 (1996).

\bibitem{Yannopapas} V. Yannopapas and A. Moroz, J. Phys: Cond. Mat. {\bf 17}, 3717 (2005). 

\bibitem{Wheeler} M.S. Wheeler, J.S. Aitchison and M. Mojahedi, Phys. Rev. B {\bf 72}, 193103 (2005).

\bibitem{LandauContMedia} L.D. Landau, E.M. Lifshitz, and L.P. Pitaevskii, {\it Electrodynamics of Continuous Media}, 2nd ed. (Elsevier, Oxford, 2007), Chap. 11.

\bibitem{Zhang} See, for example, S. Zhang, W. Fan, N.C. Panoiu, K.J. Malloy, R.M. Osgood, and S.R.J. Brueck, Phys. Rev. Lett. {\bf 95}, 137404 (2005).

\bibitem{uniaxialRC} See, for instance, L. Hu and S.T. Chui, Phys. Rev. B {\bf 66} 085108 (2002). 

\bibitem{TeltlerHenvis}This section is partially based on S. Teitler and B.W. Henvis, J. Opt. Soc. Am. {\bf 60}, 830 (1970).

\bibitem{BarashAnisotropy} Yu. S. Barash and V.L. Ginzburg, Usp. Fiz. Nauk {\bf 116}, 5 (1975), translated in Sov. Phys. Usp. {\bf 18}, 305 (1975); Y. Barash, Izv. Vyssh. Uchebn. Zaved. Radiofiz. {\bf 12}, 1637 (1978), translated in Radiophysics and Quantum Electronics {\bf 21}, 1138 (1978).  

\bibitem{Casimirtorque} J.N. Munday, D. Iannuzzi, Y. Barash, and F. Capasso, Phys. Rev. A {\bf 71}, 042102 (2005).

\bibitem{slabs} See, for instance, S.A. Ellingsen and I. Brevik, J. Phys. A {\bf 40}, 3643 (2007); I.G. Pirozhenko and A. Lambrecht, Phys. Rev. A {\bf 77}, 013811 (2008).

\bibitem{Tomas} M.S. Toma\v{s}, Phys. Rev. A {\bf 66}, 052103 (2002).

\bibitem{Tomas2} M.S. Toma\v{s}, Phys. Lett A {\bf 342}, 381 (2005).

\bibitem{Ruppin} R. Ruppin, Opt. Comm. {\bf 182}, 273 (2000).

\bibitem{ParsegianWeiss} V.A. Parsegian and G.H. Weiss, J. Adhes. {\bf 3}, 259 (1972). 

\bibitem{LeonhardtAnisotropic} T.G. Philbin and U. Leonhardt, arXiv:0806.4752, (2008).

\bibitem{KennethNussinov} O. Kenneth and S. Nussinov, Phys. Rev. D {\bf 63}, 121701(R) (2001).

\bibitem{Bruno} P. Bruno, Phys. Rev. Lett. {\bf 88}, 240401 (2002).

\bibitem{Tanaka} T. Tanaka, A. Ishikawa and S. Kawata, Phys. Rev. B {\bf 73},  125423 (2006).

\bibitem{Schilling} J. Schilling, Phys. Rev. {\bf E 74}, 046618 (2006).

\bibitem{MostPRL} F. Chen, G.L. Klimchitskaya, U. Mohideen and V.M. Mostepanenko, Phys. Rev. Lett. {\bf 90}, 160404 (2003).

\bibitem{MaxwellGarnett} J.C. Maxwell Garnett, Phil. Trans. Roy. Soc. A {\bf 203}, 385 (1904).

\bibitem{foot1} We should point out that any natural resonances that the embedding dielectric might have are already taken into account into (\ref{permittivityMG}).

\bibitem{Huang} K.C. Huang, M.L. Povinelli and J.D. Joannopoulos, Appl. Phys. Lett. {\bf 85}, 543 (2004).

\bibitem{Schuller} J.A. Schuller, R. Zia, T. Taubner and M.L. Brongersma, Phys. Rev. Lett. {\bf 99}, 107401 (2007).

\bibitem{foot2} By negative index medium we mean a material that presents a (nearly real) refractive index $n$ over a given bandwidth.  

\bibitem{MillsBurstein} D.L. Mills and E. Burstein, Rep. Prog. Phys. {\bf 37}, 817 (1977); H. Yasumoto (ed.) {\it Electromagnetic Theory and Applications for Photonic Crystals}, (CRC Taylor and Francis, Boca Raton,2006).

\bibitem{BohrenHuffman} C.F. Bohren and D.R. Huffman, {\it Absorption and Scattering of Light by Small Particles} (John Wiley and Sons, 1983), Chap. 4. 

\bibitem{foot3} This relation does not hold for arbitrarily high frequencies, but the integral tails off in this region anyway.

\bibitem{CasimirPolder} C.I. Sukenik, M.G. Boshier, D. Cho, V. Sandoghdar, and E.A. Hinds, Phys. Rev. Lett. {\bf 70}, 560 (1993); A. Landragin, J.-Y. Courtois, G. Labeyrie, N. Vansteenkiste, C.I. Westbrook and A. Aspect, Phys. Rev. Lett {\bf 77}, 1464 (1996); F. Shimizu, Phys. Rev. Lett. {\bf 86}, 987 (2001); V. Druzhinina and M. DeKieviet, Phys. Rev. Lett. {\bf 91}, 193202 (2003); T.A. Pasquini, Y. Shin, C. Sanner, M. Saba, A. Schirotzek, D.E. Pritchard and W. Ketterle, Phys. Rev. Lett. {\bf 93}, 223201 (2004).   

\bibitem{Cornell} M. Antezza, L.P. Pitaevskii and S. Stringari, Phys. Rev. A {\bf 70}, 053619 (2004); D.M. Harber, J.M. Obrecht, J.M. McGuirk and E.A. Cornell, Phys. Rev A {\bf 72}, 033610 (2005);  J.M. Obrecht, R.J. Wild, M. Antezza, L.P. Pitaevskii, S. Stringari and E.A. Cornell, Phys. Rev. Lett. {\bf 98}, 063201 (2007).

\bibitem{Buhmann} S.Y. Buhmann and D-G. Welsh, Prog. Quant. Elect., {\bf 31}, 51 (2008).


\end{thebibliography}
\end{document}